\documentclass{raa}
\usepackage{graphicx,times}             
\usepackage{natbib}
\usepackage{amssymb,amsmath}
\usepackage{verbatim}  
\usepackage{bm}  
\bibpunct{(}{)}{;}{a}{}{,}

\usepackage[pagebackref=true]{hyperref}
\hypersetup{colorlinks = true, linkcolor = green, anchorcolor = red, citecolor = blue, filecolor = red, pagecolor = red, urlcolor = red}

\begin{document}
\title{ Astronomical Site Monitoring System at Lijiang Observatory
}

\setcounter{page}{1}          

\author{
Yu-Xin Xin\inst{1,2,3,4,5}, Jin-Ming Bai\inst{1,3,4}, Bao-Li Lun\inst{1,3,4}, Yu-Feng Fan\inst{1,3,4}, Chuan-Jun Wang\inst{1,3,4}, Xiao-Wei Liu\inst{5}, Xiao-Guang Yu\inst{1,3,4}, Kai Ye\inst{1,3,4}, Teng-Fei Song\inst{1,3,4}, Liang Chang\inst{1,2,3,4}, Shou-Sheng He\inst{1,3,4}, Ji-Rong Mao\inst{1,3,4}, Liang Xu\inst{1,3,4}, Ding-Rong Xiong\inst{1,3,4}, Xi-Liang Zhang\inst{1,3,4}, Jian-Guo Wang\inst{1,3,4}, Xu Ding\inst{1,2,3,4}, Hai-Cheng Feng\inst{1,2,3,4}, Xiang-Kun Liu\inst{5}, Yang Huang\inst{5}, Bing-Qiu Chen\inst{5}
}

   \institute{ Yunnan Observatories, Chinese Academy of Sciences,Kunming 650216, China; {\it xyx@ynao.ac.cn; baijinming@ynao.ac.cn}\\
    \and
     University of Chinese Academy of Sciences, Beijing 100049,China\\
    \and
    Key Laboratory for the Structure and Evolution of Celestial Objects, Chinese Academy of Sciences, Kunming 650216, China\\
    \and
    Center for Astronomical Mega-Sciences, Chinese Academy of Sciences,Beijing 100012,China\\
    \and
    South-Western Institute for Astronomy Research, Yunnan University, Kunming 650500, China\\
\vs \no
   {\small Received 2020 Feb 20; accepted 2020 April 21}
   }

\abstract{ We installed two sets of Astronomical Site Monitoring Systems(ASMSs) at Lijiang Observatory(GMG), for the running of the 2.4-meter Lijiang optical telescope(LJT) and the 1.6-meter Multi-channel Photometric Survey Telescope (Mephisto). The Mephisto is under construction. The ASMS has been running on robotic mode since 2017. The core instruments: Cloud Sensor, All-Sky Camera and Autonomous-DIMM that are developed by our group, together with the commercial Meteorological Station and Sky Quality Meter, are combined into the astronomical optical site monitoring system. The new Cloud Sensor's Cloud-Clear Relationship is presented for the first time, which is used to calculate the All-Sky cloud cover. We designed the Autonomous-DIMM located on a tower, with the same height as LJT. The seeing data have been observed for a full year. ASMS's data for the year 2019 are also analysed in detail, which are valuable to observers.
}
\authorrunning{Y.-X. Xin et al. }            
\titlerunning{Astronomical Site Monitoring System at Lijiang Observatory}  
\maketitle

%
\section{Introduction}           
\label{sect:intro}

\begin{figure}
   \centering
   \includegraphics[width=0.8\textwidth, angle=0]{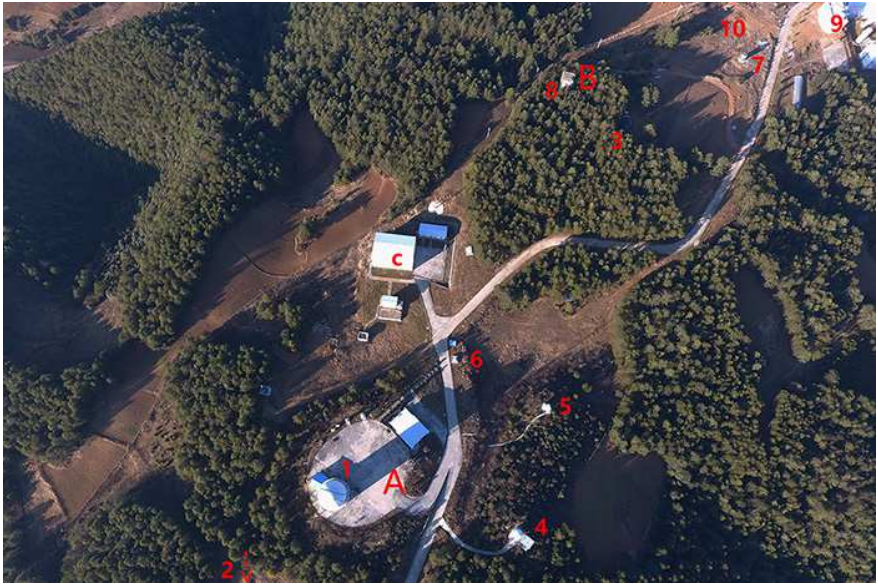}
   \captionsetup{justification=centering}  
   \caption{\small Lijiang Observatory  (A: ASMS-A on the ground; B: ASMS-B on the top of water tower; c: 3.2m Vacuum Coating mechine room; $1\sim10$: refer to Table.\ref{table:telescopes})}
   \label{fig:LJO}
\end{figure}

 Lijiang Observatory(Fig.\ref{fig:LJO}) ($3200$m, E: $100^{\circ} 01'48''$, N:$26^{\circ}41'42''$, IAU code: $044$) is the biggest optical astronomical site in Southern China, which are located on the top of Tiejia Mountain, nearby the Gaomeigu village, Yulong County, $40$ km far from Lijiang City. "Gao-Mei-Gu" is a Naxi minority language, with the meaning of "a beautiful highland for looking at the stars". It is the best site for the 2.4-meter telescope in the site-testing campaign during the end of twentieth Century ($1993\sim2002$). The Lijiang 2.4-meter telescope (LJT) is the major optical telescope in this site, the biggest normal optical telescope in China by now. LJT was built in 2005, and it was settled down in 2007. It has opened to the world-wide astronomers since 2008 (\citealt{2m4}). 

The Astronomical Site Monitoring System (ASMS) is a pre-quisite for any observatories planning robotic and optimized scheduling of large astronomical facilities, and it is a must for any modern observatory (\citealt{VLT-ASM}; \citealt{ASM+2002}). There are 10 telescopes (Table \ref{table:telescopes}) performing in the Lijiang Observatory(GMG) by the end of 2019, four of them are running on Robotic Mode (\citealt{bootes4}), which seriously rely on the stable real-time observation condition information from ASMS. The history of ASMS development at Lijiang Observatory can be summarized as follows:\\
$\bullet$ \textbf{Oct, 2008 $\sim$ May, 2011}: (1.5 meters high) Built a simple weather station for six parameters\footnote{Temperature,Relative Humidity,Wind Speed,Wind Direction,Pressure,Rain.}.\\
$\bullet$ \textbf{Jan, 2012 $\sim$ Oct, 2016}: (6 meters high) Installed a Davis Weather Station for six parameters.\\
$\bullet$ \textbf{Nov, 2013 $\sim$ now}: (2 meters high) Installed an Allsky Camera.\\
$\bullet$ \textbf{Nov, 2016 $\sim$ now}: (6 meters high, ASMS-A) Installed Davis Wireless Weather Station for eight parameters\footnote{Temperature,Relative Humidity,Wind Speed,Wind Direction,Pressure,Rain,UV index, Solar Radiation.}, single-zone CloudSensor, All sky Camera, SQM-LE and Remote-DIMM(RDIMM).\\
$\bullet$ \textbf{Jan, 2019 $\sim$ now}: (14 meters high, ASMS-B) Installed Davis Wireless Weather Station for eight parameters, double-zone CloudSensor, Allsky Camera, SQM-LE and Autonomous-DIMM(ADIMM).

\begin{table}
\bc
\begin{minipage}[]{60mm}
\caption[]{Telescopes in GMG \label{table:telescopes}}\end{minipage}
\setlength{\tabcolsep}{1pt}
\small
 \begin{tabular} {p{1cm} p{2cm} p{2cm} p {2cm} p{8cm} }
  \hline\noalign{\smallskip}
Num & Name & Aperture/m & Mode & Introduction \\
  \hline\noalign{\smallskip}
 1 & LJT & 2.4 & Remote & Lijiang 2.4m Telescope (R-C Cassegrain and Nasmyth Focus)  \\
 2 & 1m8 & 1.8 & Remote & Lijiang 1.8m Telescope (R-C Coude Focus) \\
 3 & Mephisto & 1.6 & Robotic & Multi-channel Photometric Survey Telescope (FOV: $2^{\circ}*2^{\circ}$)  \\
 4 & STT & 0.7 & Remote & Sino-Thai Telescope (Nasmyth Focus)  \\
 5 & BOOTES-4 & 0.6 & Robotic & Burst Optical Observer and Transient Exploring System 4\#  \\
 6 & TAT-3 & 0.1 & Robotic & Taiwan Automatic Telescope network 3\# \\
 7 & TEST & 0.45 & Remote & Yunnan-HongKong Transiting Exoplanet Survey Telescope \\
 8 & ADIMM & 0.3 & Robotic & Automated Differential Image Motion Monitor \\
 9 & LCT & 0.1 & Remote & Lijiang Coronagraph Telescope \\
 10 & FASOT-1B & 0.4 & Remote & Fiber-Array Solar Telescope \\
\noalign{\smallskip}\hline
\tablecomments{0.76\textwidth}{3\# Mephisto Telescope's construction will be completed by the end of 2020}
 \end{tabular}
 \ec
\end{table}

There are four subsystems in ASMS: the Meteorological Station(in Sect. \ref{subsect:MS}), All sky-Information Acquisition System(in Sect. \ref{subsect:AIAS}), ADIMM system (in Sect. \ref{subsect:ADIMM}) and Video Surveillance System (in Sect. \ref{subsect:VSS}). We present the detailed designs for the Cloud Sensor, Allsky Camera and ADIMM system. In Sect. \ref{sect:annual_data_analysis}, we analyzed the ASMS's dataset based on the full year of 2019, obtained the basic observation condition of Lijiang Observatory, including the weather condition (from Sect. \ref{subsect:Temp} to Sect. \ref{subsect:UV}), darkness condition(in Sect.  \ref{subsect:darkness}), observable times (in Sect.  \ref{subsect:obs_hours}) and seeing condition (in Sect. \ref{subsect:seeing}). In the last part Sect. \ref{sect:conclusion}, we give the Conclusion of this work.

\section{Astronomical Site Monitoring System Overview}
\label{sect:overview}
A modern Astronomical Site Monitoring System provides the following indispensable information for the observatory : (1) Real-time collecting all the essential Astroclimate information for Observatory Control System (OCS) and Robotic Autonomous Observatory (RAO)(\citealt{RAO}), such as Cloud Cover, Wind Speed, Humidity, Dust, Temperature, Dew Point and Rain.
(2) Collecting all the auxiliary site information for science data reduction and station assessment, such as seeing, sky brightness, wind direction, air pressure, UV index and solar radiation.
(3) Monitoring the status of the sky and the astronomical facilities, such as dome, telescope and science instruments, through All sky-images and monitoring videos.
All information is also used in other ways: (1) Optimization of observation strategy. (2) Improving the dome seeing, mirror seeing and instruments seeing. (3) Assist for Adaptive Optics (AO) system. (4) Build the basis for site weather prediction. (\citealt{VLT-ASM}; \citealt{Tokovinin2002}; \citealt{SAAO2012})

\begin{figure}[h]
  \begin{minipage}[t]{0.5\linewidth}
  \centering
   \includegraphics[width=50mm,height=65mm]{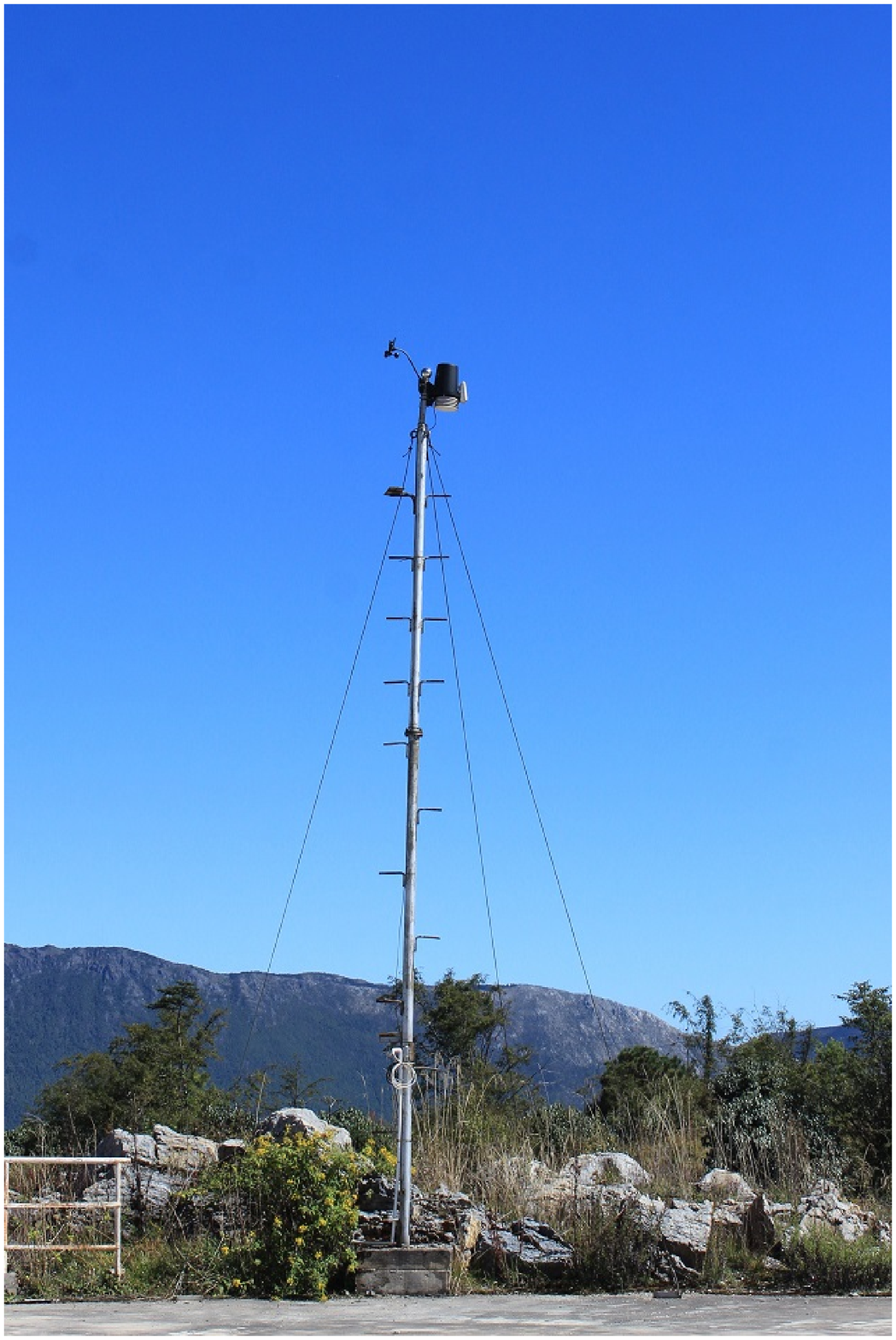}
   \caption{{\small Meteorological Station of ASMS-A} }
   \label{fig:asms-a}
  \end{minipage}
  \begin{minipage}[t]{0.5\textwidth}
  \centering
   \includegraphics[width=50mm,height=65mm]{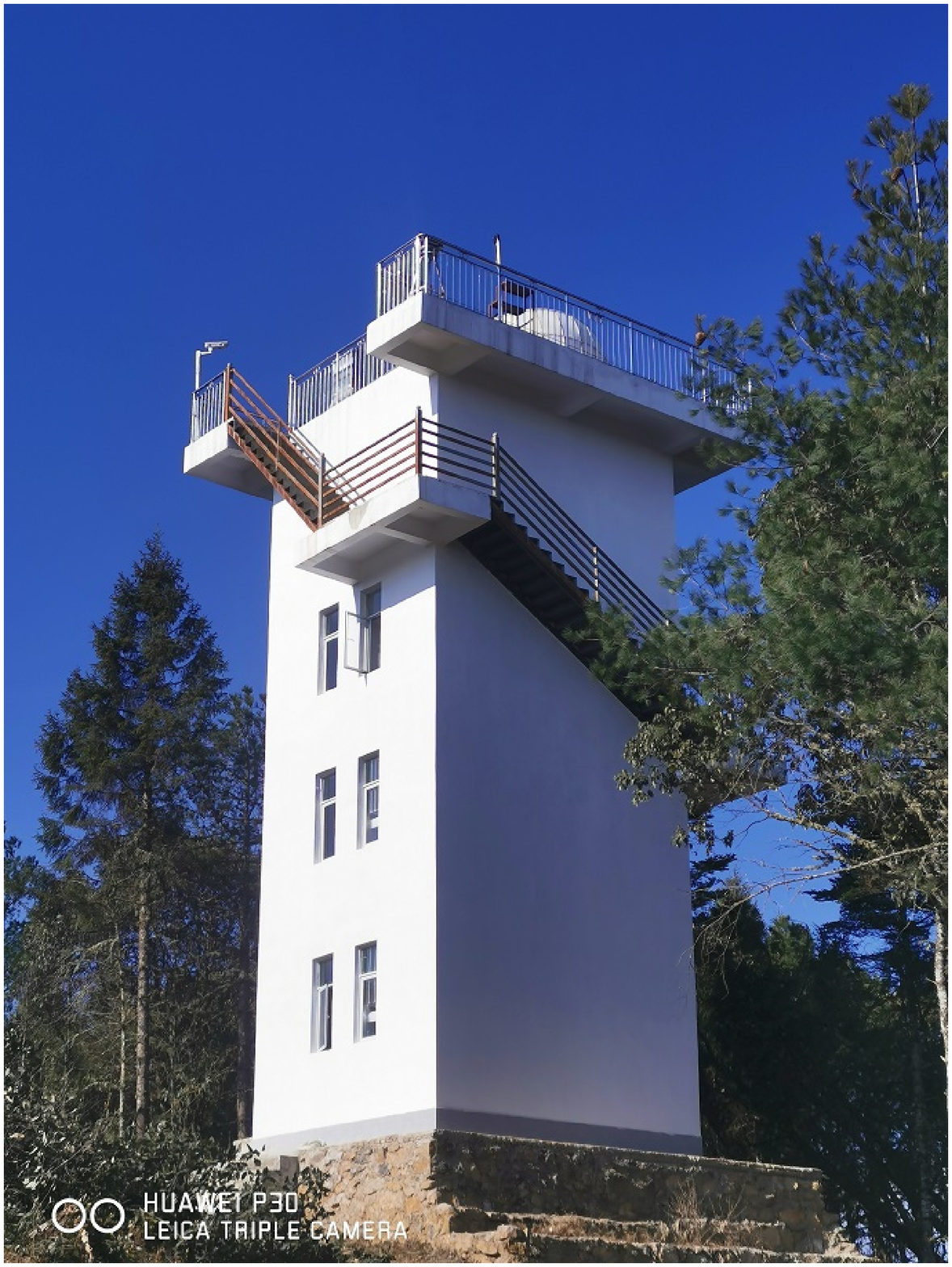}
  \caption{{\small Water Tower of ASMS-B}}
  \label{fig:watertower}
  \end{minipage}
  \label{fig:asms}
\end{figure}

\begin{figure}
   \centering
   \includegraphics[width=0.85\textwidth]{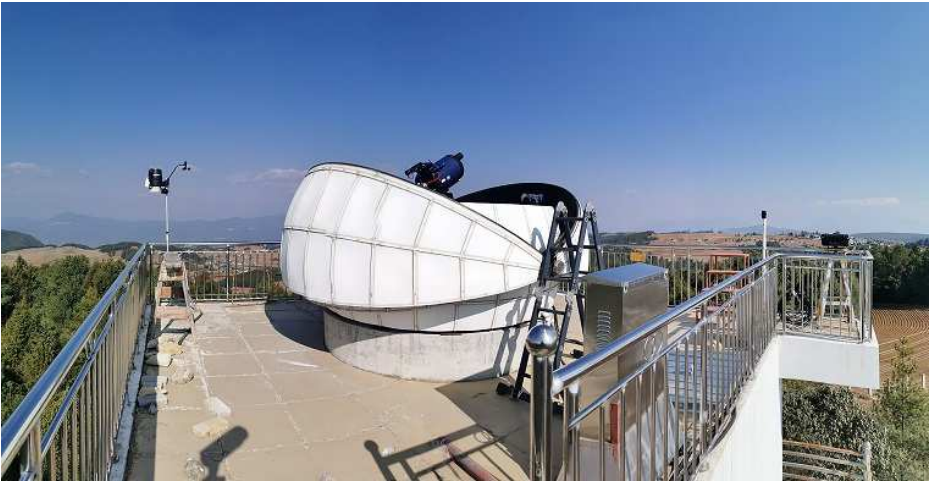}
   \captionsetup{justification=centering}  
   \caption{\small ASMS-B Layout}
   \label{fig:asms-b}
\end{figure}

We developed two sets of ASMSs in Lijiang Observatory, the ASMS-A (Fig. \ref{fig:asms-a}) located on the square around LJT, and ASMS-B (Fig.\ref{fig:watertower} and Fig.\ref{fig:asms-b}) located on a Water Tower near the site of Mephisto Telescope. Three parts are included in each ASMS: a Meteorological Station, All sky-Information Acquisition System and DIMM system. The Dust Meter is not needed for GMG, there are neither La Palma's Calima nor dust in Xinglong Observatory(\citealt{xinglong+2015}), except a small amount of potato pollen during July and August.

\subsection{Meteorological Station}
\label{subsect:MS}
A calibrated weather station that was developed by our group was used at GMG from October 2008, and it was replaced by a calibrated commercial weather station--Davis Vantage Pro2 in February 2012. It was replaced one again by a wireless version in November 2016. The mounting height is 6 meters above the ground, and we installed the other Davis Vantage Pro2 on a Water Tower 14-meters high in January 2019. All the Davis Stations have the same performance parameter(see Table \ref{table:davis}), and the dew point is calculated value depending on air temperature and humidity, according to the formula equation\textasciitilde(\ref{eq:dewpoint01}) and equation\textasciitilde(\ref{eq:dewpoint02}). The meteorological data are read directly from the console per minute through our python program, and all the data are stored in the MySQL database.

\begin{table}
\centering
\caption{Technical parameters of meteorological station}
 \label{table:davis}
 \begin{tabular}{p{3cm} p{3cm} p{1.5cm} p{1.5cm}}
  \hline\noalign{\smallskip}
Parameter & Scope & Resolution & Precision  \\
  \hline\noalign{\smallskip}
Temperature & $-40^{\circ}$C$\sim65^{\circ}$C & 1 $^{\circ}$C & $\pm0.5$ $^{\circ}$C \\
Relative Humidity & $0 \%\sim100$ \% & 1 \% & 3 \% \\
Wind Speed & $1\sim67$ m s$^{-1}$ & 0.1 m s$^{-1}$ & $\pm5$\% \\
Wind Direction & $0\sim360$ & 1 $^{\circ}$ & 4 $^{\circ}$ \\
Air Pressure & $0\sim1\,080$ hPa & 0.1 hPa & 1.0 hPa \\
Rain & $0\sim9\,999$ mm & 0.2 mm & $\pm4$\% \\
Solar Radiation & $0\sim1\,800$ W m$^{-2}$ & 1 W m$^{-2}$  & 5 \% \\
UV Index & $0\sim16$ Index & 0.1 Index & 5 \% \\
  \hline\noalign{\smallskip}
 \end{tabular}
\end{table}

The dew point is the temperature at which air is saturated with water vapor, which is the gaseous state of water. The roof has to be closed when the dew point is near the air temperature.
Temperature and relative humidity are used for calculating the dew point, according to the Magnus-Tetens formula. 
It allows us to obtain the accurate results (with an uncertainty of $0.35^{\circ}$C) for the temperature with the range from $-45^{\circ}$C to $60^{\circ}$C.

\begin{equation}
\alpha(T,RH)=\ln(RH/100)+ a \cdot T / (b+T)\vspace{1ex}
\label{eq:dewpoint01}
\end{equation}

\begin{equation}
T_{s}=(b\cdot\alpha(T,RH) / (a- \alpha(T,RH)))\vspace{1ex}
\label{eq:dewpoint02}
\end{equation}

where $T_{s}$ is the dew point, $T$ is the temperature, RH is the relative humidity of the air, $a$ and $b$ are coefficients, $a=17.62$ and $b=243.12$. Temperature, Humidity and Pressure are also used for the atmospheric refraction correction.

\subsection{Allsky-Information Acquisition System}
\label{subsect:AIAS}
\begin{figure}
   \centering
   \includegraphics[width=0.7\textwidth]{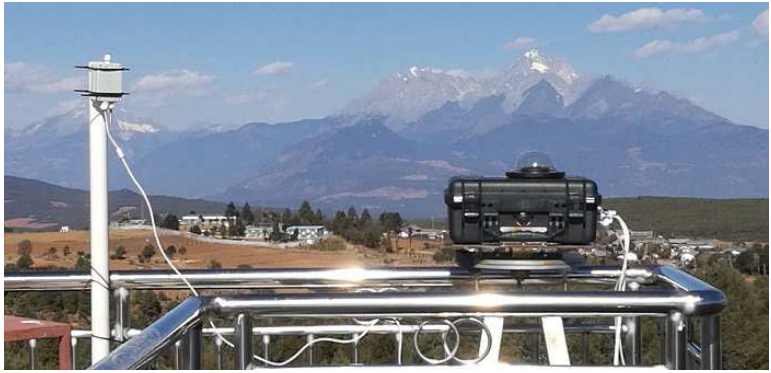}
   \captionsetup{justification=centering}  
   \caption{\small All sky-Information Acquisition System on top of Water Tower  ( Left: Cloud Sensor I; Right: All sky Camera and SQM-LE inside the black box; Background: Haba Snow Mountain(left: 70km far away, 5396m high) and Jade Dragon Snow Mountain (right: 50km far away, 5596m high))}
   \label{fig:aias}
   \end{figure}

The All-sky cloud cover (values range from $0$ to $10$: $0$ represents $0\%$ cloudiness, $10$ represents $100\%$ cloudiness ) is critical data for remote or robotic telescopes, and it is more important than any other astro-climate dataset for a modern observatory. In most cases the roof has to be closed when the cloud cover is greater than 6 (60\% cloudiness), for preventing the potential rain in the clouds. All-sky images are still necessary for us to know the distribution and thickness of cloud. We developed the All sky-Information Acquisition System (Fig. \ref{fig:aias}) for getting the all-sky cloud cover, the all-sky image, and the night darkness per minute, and it has been realized the real-time acquisition of all-sky information since 2016.

\subsubsection{Cloud Sensor}
\label{subsubsect:cloud}

Cloud detection prevents low quality observations.
We designed two new Cloud Sensors using the single and dual zone infra-red thermometer MLX90614 ( $0.5^{\circ}$C accuracy from $-70^{\circ}$ to $380^{\circ}$ ). The sensor is also used by the Aurora Cloud Sensor\footnote{http://auroraeurotech.com/}. The design principle is that Cloud Cover is inversely proportional to the temperature difference between the sky and the ground ambient. The new Cloud Sensor invention patents (CN201711497699.7) is under review.

Cloud Sensor I (Fig. \ref{fig:aias}): using one dual-zone infra-red thermometer to measure the average infrared temperature $T_{\rm sky}$ for the sky above 30 degrees level, and measure the ambient temperature $T_{\rm amb}$ using the digital thermometer DS18B20 ( $\pm0.5$ $^{\circ}$C accuracy from $-10^{\circ}$ to $85^{\circ}$ ), and then using the Cloud-Clear Formula (Eq. \ref{eq:ccf01} and Eq. \ref{eq:ccf02}) to calculate the cloud cover $C_{\rm cloud}$  of the 120 degree range of the sky.

\begin{figure}[h]
  \begin{minipage}[t]{0.5\linewidth}
  \centering
   \includegraphics[width=65mm,height=45mm]{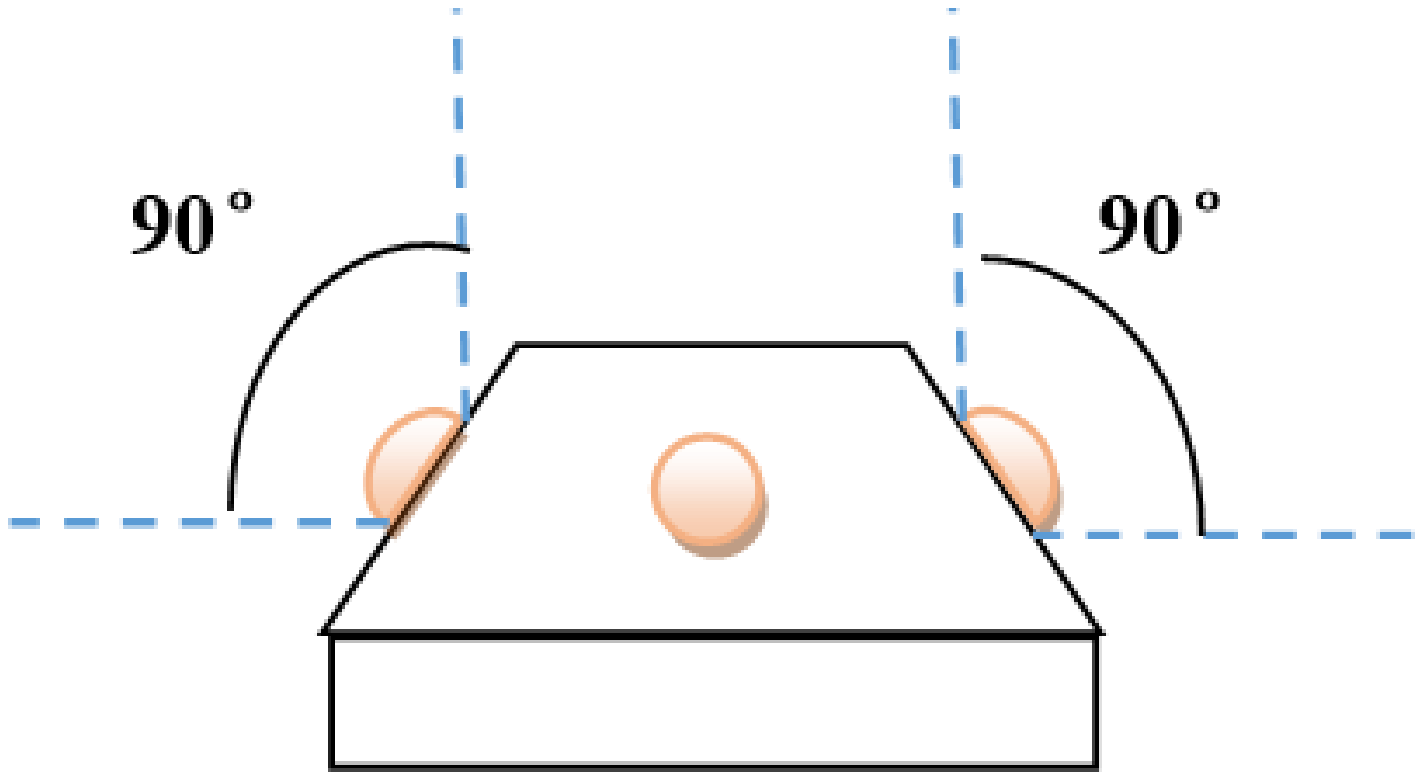}
   \caption{{\small Cloud Sensor II} }
   \label{fig:cloudsensor}
  \end{minipage}
  \begin{minipage}[t]{0.5\textwidth}
  \centering
   \includegraphics[width=65mm,height=45mm]{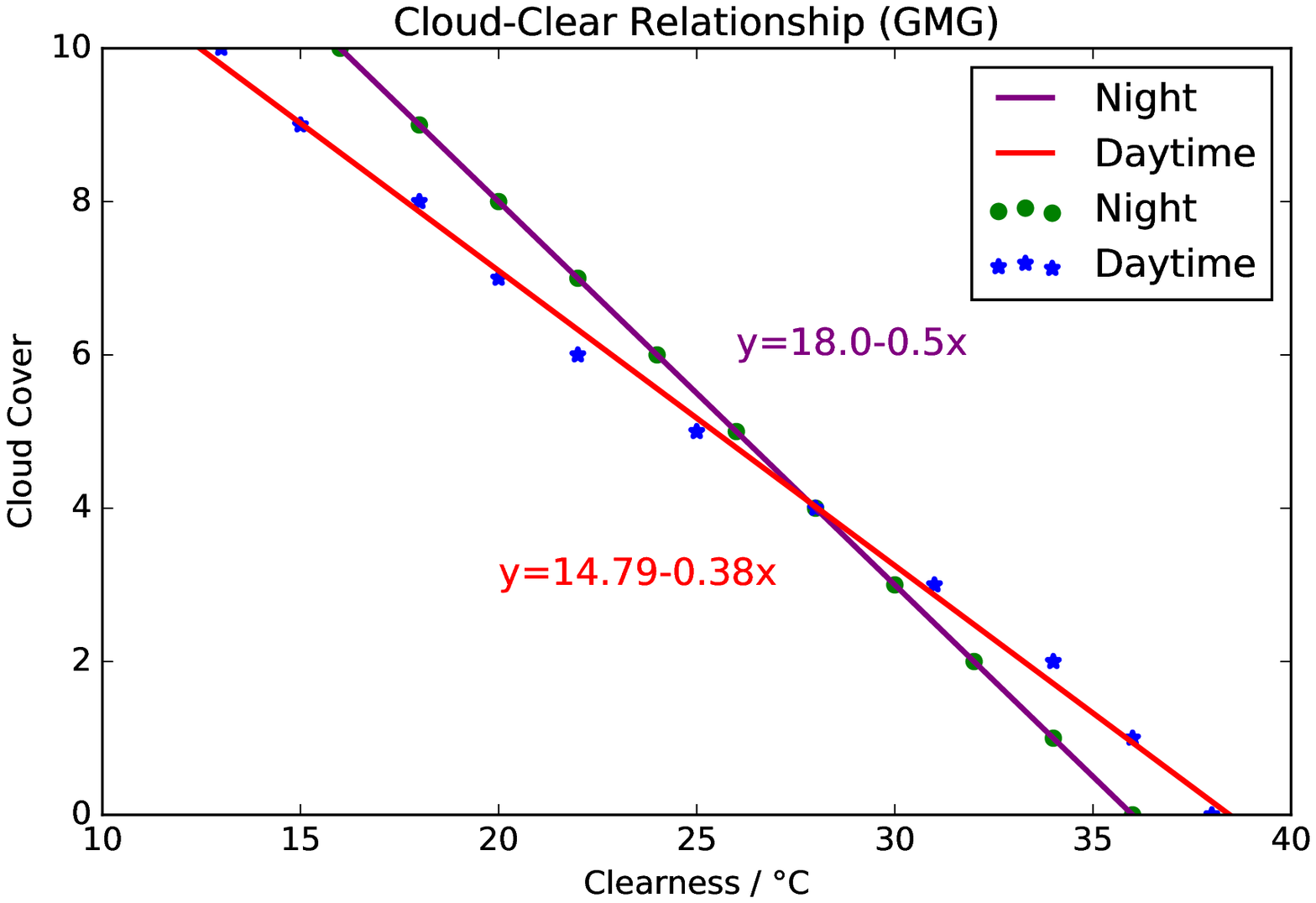}
  \caption{{\small Cloud Clear Relationship for GMG }}
  \label{fig:ccr}
  \end{minipage}
  \label{fig:cc}
\end{figure}

Cloud Sensor II (Fig. \ref{fig:cloudsensor}): using four single-zone infra-red thermometers to separately measure the four average infrared temperatures $T_{\rm eastsky}$, $T_{\rm southsky}$, $T_{\rm westsky}$ and $T_{\rm northsky}$ in four direction scope of the all-sky , and measure the ambient temperature $T_{\rm amb}$ using the digital thermometer DS18B20, and then using the Cloud-Clear Formula to calculate the cloud cover of all over the sky.

\begin{figure}[h]
  \begin{minipage}[t]{0.5\textwidth}
  \centering
   \includegraphics[width=80mm,height=50mm]{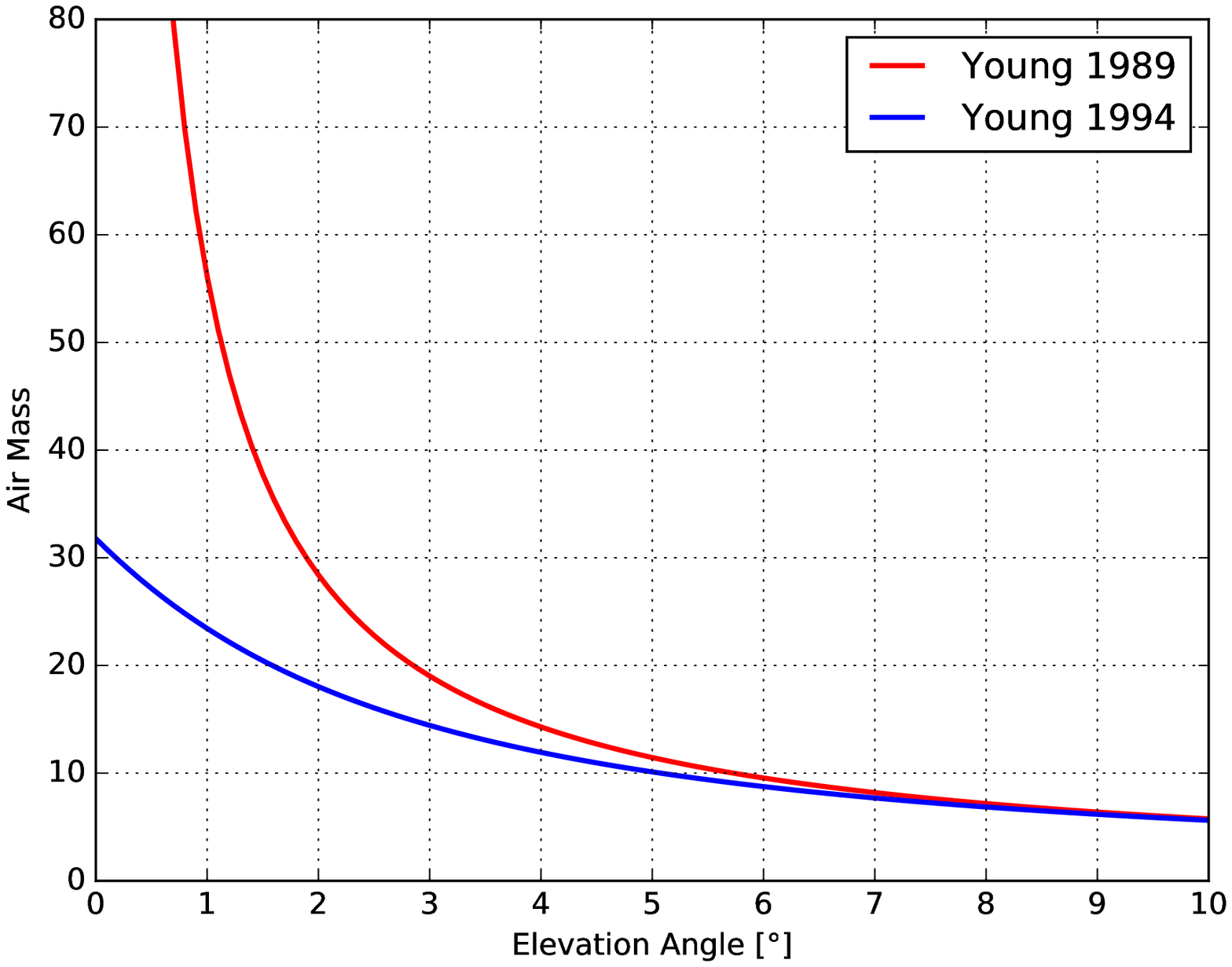}
  \caption{{\small Air Mass at $h=0^{\circ}\sim10^{\circ}$}}
  \label{fig:airmass01}
  \end{minipage}
    \begin{minipage}[t]{0.5\linewidth}
  \centering
   \includegraphics[width=80mm,height=50mm]{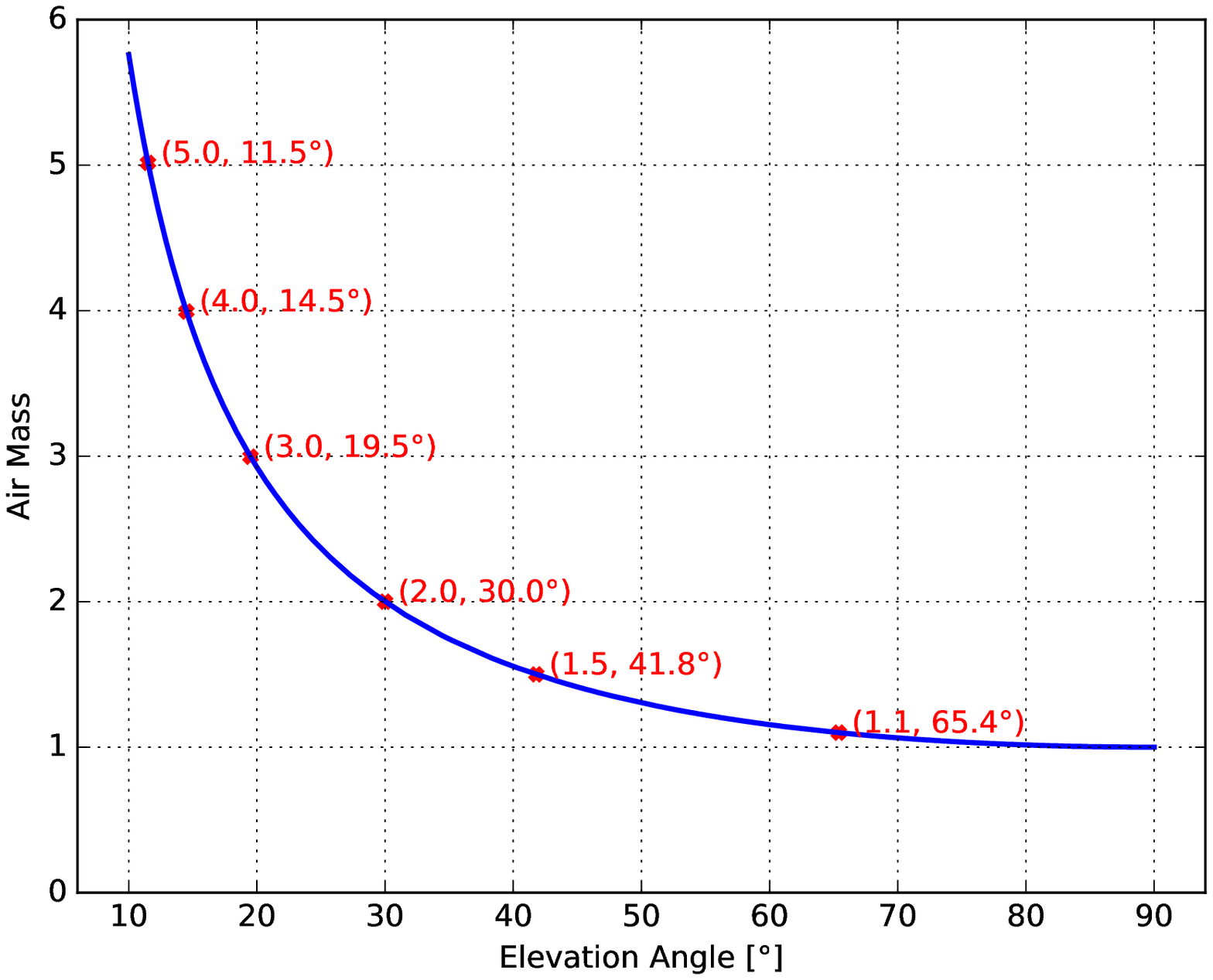}
   \caption{{\small Air Mass at $h=10^{\circ}\sim90^{\circ}$} }
   \label{fig:airmass02}
  \end{minipage}
\end{figure}

The Cloud Sensor I to collect 24 hours of cloud cover in Lijiang Observatory's ASMS has been used since Nov 2016, and
the sampling frequency is one minute, detecting the sky from $30$ to $90$ degrees. The 2.4-meter telescope has the observing height with the range of $20\sim90$ degrees. Usually the astronomical observation has the height above $30$ degrees, where the airmass is 2 (see Fig. \ref{fig:airmass01}). We will introduce the airmass in later section.

The coefficient $(b, k)$ in the Cloud-Clear Formula is independent on the local topographic, the climatic conditions, the sea level, and the altitude of the sun. We combine the cloud cover in all-sky camera image and $C_{\rm clear}$ data and use the least square method to get the coefficient for the daytime and nighttime Cloud-Clear Formula (Fig. \ref{fig:ccr}).

\begin{equation}
C_{clear}=T_{amb}-T_{sky}
\label{eq:ccf01}
\end{equation}
\begin{equation}
C_{cloud}=b-k \cdot C_{clear},
\begin{cases}
C_{cloud}=0, & \mbox {if } C_{clear} \mbox{$\ge$ upper limit} \\
C_{cloud}=10, & \mbox {if } C_{clear} \mbox{$\leq$ lower limit}
\end{cases}
\label{eq:ccf02}
\end{equation}

$\bullet$ \textbf{Air Mass}

If star is at the zenith (altitude $90^{\circ}$ degrees above the horizon), the light goes through the air mass 1.
The calculations of the air mass $X(z)$ take different approximation(\citealt{airmass01}; \citealt{airmass03}) at different alt angles, and we have the references to equation\textasciitilde(\ref{eq:airmass01}) for the height from 10 to 90 degrees (Fig. \ref{fig:airmass02}) and equation\textasciitilde(\ref{eq:airmass03}) for the height from 0 to 10 degrees, to avoid the air mass to be infinite at 0 degree (red line in Fig. \ref{fig:airmass01}). Actually the starlight is bent on its path through the atmosphere due to the refraction, so that the real change curve of the airmass is the blue line in Figure \ref{fig:airmass01}. Otherwise we would never see the sun rise or sun set.

Seeing (spatial resolution), Sky brightness (limiting magnitude) and Atmospheric extinction (transparency) are three major effects of the air conditions on the ground-based optical telescopes. They are all directly related to the air mass, namely thickness of the atmosphere.

\begin{equation}
X(z)=\frac{1}{sin(h)}=\frac{1}{cos(z)}, \mbox{\rm h} \mbox{$\in$[10,90]}, \mbox{\rm z} \mbox{$\in$[0,80]}
\label{eq:airmass01}
\end{equation}

\begin{equation}
\begin{aligned}
X(z)=\frac{1.003198cos(z)+0.101632}{cos^{2}(z)+0.09056cos(z)+0.003198},\mbox{\rm z} \mbox{$\in$[80,90]}\\
\end{aligned}
\label{eq:airmass03}
\end{equation}

\subsubsection{All-Sky Camera}

Several versions of All-Sky Cameras have been developed since 2013 (see Table \ref{table:allskycamera}). We introduced the temperature and humidity control unit, double transparent shield, electronic shutter replacement the mechanical shutter, anti-ghost measures, automatic control and acquisition procedures based on Linux for the All-Sky Camera, and we have solved the following issues:(1) Clear hemispherical acrylic hood's frosting and condensation of moisture issues.(2) Shutter life issues. (3) Ghost issue.(4) Ultra-low temperature environment flexibility. (5) Robotic and high frequency sampling. The all-sky camera is running on robotic mode, and one image per minute is on auto exposure mode. We have six-year all-sky images in the GMG station

\begin{table}
\caption[]{All-sky Cameras \label{table:allskycamera}}
\setlength{\tabcolsep}{1pt}
\small
 \begin{tabular} {p{1.5cm} p{3cm} p{2cm} p {4cm} p{2cm} p{3.5cm}  }
  \hline\noalign{\smallskip}
Version & Lens & Camera & Enclosure & Working Temperature & Location of use \\
  \hline\noalign{\smallskip}
V2013 & Sigma 4.5mm F2.8 & Canon 600D & Modified surveillance shields & $-10^{\circ} \sim 30^{\circ}$ & GMG \\
V2014 & QHY fish-eye lens & QHY5II & Metal Box & $-10^{\circ} \sim 30^{\circ}$ & GMG \\
V2015 & Sigma 4.5mm F2.8 & Canon 600D & PELI Plastic Box & $-45^{\circ} \sim 40^{\circ}$ & GMG, Zhongshan Station \\
V2016 & Sigma 4.5mm F2.8 & ASI1600MC & PELI Plastic Box & $-30^{\circ} \sim 40^{\circ}$ & GMG, Iceland Station\\
V2018 & Sigma 4.5mm F2.8 & ASI1600Pro & Metal Box and Wood Box & $-80^{\circ} \sim 30^{\circ}$ & Dome A, Taishan Station\\
V2019 & Sigma 4.5mm F2.8 & ASI1600MC & PELI Plastic Box & $-45^{\circ} \sim 40^{\circ}$ & GMG, Huanghe Station, PondInlet Station \\
\noalign{\smallskip}\hline
 \end{tabular}
\end{table}

\subsubsection{Sky Brightness Meter}

The night sky is not absolutely dark, and it is immersed within many illumination sources, such as: the moon, the polar aurora, airglow, starlight(including the Milky Way and Orbiters), zodiacal light, gegenschein\footnote{Gegenschein,also called Counterglow, oval patch of faint luminosity exactly opposite to the Sun in the night sky.} and the artificial light pollution (sodium lamp, mercury lamp, etc). The auroras are confined to high latitudes (Antarctic and Arctic regions). The zodiacal light arises from sunlight scattered by the interplanetary dust in the solar system(\citealt{airglow}). The airglow is the luminescence of the earth's atmosphere itself, it is the main component of the moonless sky brightness. The typical moonless sky brightness value on astronomical site is $21\sim22$ mag arcsec$^{-2}$

\begin{figure}
  \begin{minipage}[t]{0.5\linewidth}
  \centering
   \includegraphics[width=30mm,height=50mm]{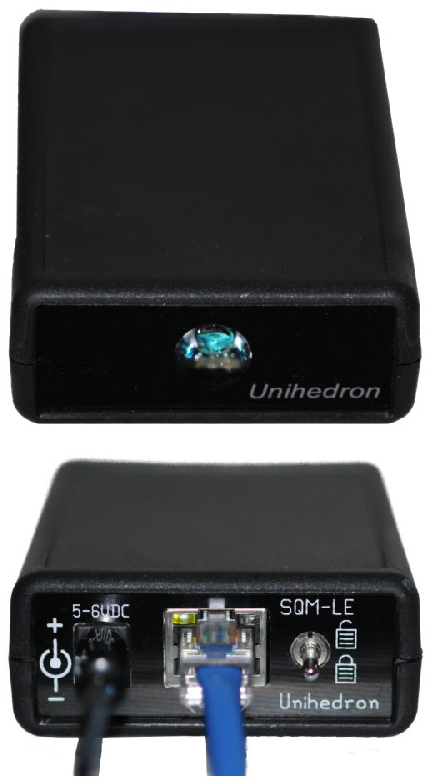}
   \caption{{\small SQM-LE } }
   \label{fig:sqmle01}
  \end{minipage}
  \begin{minipage}[t]{0.5\textwidth}
  \centering
   \includegraphics[width=70mm,height=50mm]{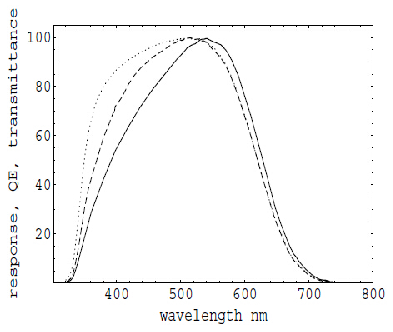}
  \caption{{\small SQM-LE normalized spectral responsivity (solid line), quantum efficiency (dashed line) and filter transmittance (dotted line). (\citealt{SQM})}}
  \label{fig:sqmle02}
  \end{minipage}
  \label{fig:sqm}
\end{figure}

Sky Quality Meter (Fig.\ref{fig:sqmle01}) is a world-wide used sensor to measure the Sky Brightness, and it becomes a world standard for assessing the darkness of a astronomical site by International Dark-Sky Association. Lijiang observatory used the SQM-LE, which was installed inside the All-Sky Information Acquisition System, pointing to the zenith through a flat glass. We have calibrated the result offset bring by the flat glass under different dark condition, and it brings 0.08 offset to the value from SQM-LE. The central wavelength of normalized spectral responsivity is about $550nm$ (Fig.\ref{fig:sqmle02}), corresponding the Johnson V-band, so that the final result is the zenith V-band sky brightness level.

\subsection{ADIMM System}
\label{subsect:ADIMM}
Ground-based astronomical optical telescopes' theoretical diffraction limit angle is $\theta \approx 1.22\lambda/d$, which is actually the resolution $\theta$ is limited by the coherent diameter $r_{0}$ of atmospheric turbulence is $\theta=(4/\pi)\lambda/r_{0} (d>>r_0)$, which is based on the Kolmogorov Model(\citealt{Roddier1981}). Therefore the atmospheric seeing is a critical parameter for any Observatory in the world. For example: a 2-meter telescope that is located at $1''$ seeing site is equivalent to a 4-meter telescope that is located at $2''$  seeing site, they can get the same Signal-to-Noise Ratio under the same exposure time, and the former have the better spatial resolution, moreover the AO system adjustable units are much less at good seeing site. equation.\ref{Eq:AO_units} gives the relationship between Fried parameter $r_0$ (Seeing) and the number(N) of actuators of the AO system on a telescope with the aperture is D. It is well known that $r_0(\lambda) \propto \lambda^{6/5}$, so the AO system is simple in infrared wavelength than the visible wavelength.

\begin{equation}
\label{Eq:AO_units}
    N(\lambda)=(\frac{D}{r_0(\lambda)})^2
\end{equation}

The FWHM of the corresponding point spread function was estimated numerically in a function of wavelength as equation. \ref{Eq:FWHM} (\citealt{FWHM1988}), and it is seeing which is called by astronomers, the unit is arc-second. The telescope's image quality is directly related to the seeing and its essence is the Fried parameter $r_{0}$.

\begin{equation}
\label{Eq:FWHM}
    FWHM=0.98\lambda/r_{0}
\end{equation}

There are several methods in history to assess the atmospheric seeing. Star trail photographs have been used to assess image quality(\citealt{Harlan1965}), and photoelectric image motion monitors can get the absolute seeing disk sizes(\citealt{Irwin1966}). Interferometers have also been used to measure the wavefront coherence(\citealt{Roddier1986}). However, all the above issues are sensitive to tracking errors and instrument vibration. The European Southern Observatory developed a differential image motion monitor (DIMM)(\citealt{ESO1990}) for Very Large Telescope site testing campaign, which measured wavefront slope differences over two small pupils some distance apart. As a differential method, DIMM is insensitive to tracking errors and instrument vibration errors.
The traditional formulae of DIMM are equation\textasciitilde(\ref{Eq_l}) and equation\textasciitilde(\ref{Eq_t}), where $d$ is the distance of two pupils, $D$ is the diameter of pupil, $\lambda=550nm$, $\sigma_l^2$, $\sigma_t^2$ are parallel and perpendicular variances on the direction of  pupil connection.
Some DIMM systems are still using the traditional formulae in the world. Considering the type of tilt, the new formulae (eq.\textasciitilde(\ref{Eq_l_2002}), eq.\textasciitilde(\ref{Eq_t_2002})) give an excellent match to the exact coefficients computed by other authors(\citealt{Martin1987}; \citealt{Tyler1994}; \citealt{Tokovinin2002}). The new formulae strongly suggest  adopting $\lambda=500nm$ as the standard for seeing data computation.

$r_{0l}$, $r_{0t}$ are parallel and perpendicular Fried parameters on the direction of  pupil connection. According to the Kolmogorov model, the Fried parameters are isotropic, so we can get the general $r_0$ equations (Eq.\ref{Eq_r0_eso} and Eq.\ref{Eq_r0_2002}), for traditional formulae and new formulae of the DIMM system.
Convert to the zenith for $r_{0}$ is equation\textasciitilde(\ref{Eq_zenith}), which is suitable for zenith distance less than $30^{\circ}$.

\begin{equation}
\label{Eq_l}
  \sigma_l^2=2\lambda^2r_{0l}^{-5/3}[0.179D^{-1/3}-0.097d^{-1/3}]
\end{equation}
\begin{equation}
\label{Eq_t}
  \sigma_t^2=2\lambda^2r_{0t}^{-5/3}[0.179D^{-1/3}-0.145d^{-1/3}]
\end{equation}

\begin{equation}
\label{Eq_l_2002}
  \sigma_l^2=0.364\lambda^2r_{0l}^{-5/3}D^{-1/3}[1-0.532(d/D)^{-1/3}-0.024(d/D)^{-7/3}]
\end{equation}
\begin{equation}
\label{Eq_t_2002}
  \sigma_t^2=0.364\lambda^2r_{0t}^{-5/3}D^{-1/3}[1-0.798(d/D)^{-1/3}+0.018(d/D)^{-7/3}]
\end{equation}

\begin{equation}
\label{Eq_r0_eso}
  r_0=\{\frac{2\lambda^2[0.358D^{-1/3}-0.242d^{-1/3}]}{(\sigma_l^2+\sigma_t^2)}\}^{3/5}
\end{equation}
\begin{equation}
\label{Eq_r0_2002}
  r_0=\{\frac{0.364\lambda^2D^{-1/3}[2-1.33(d/D)^{-1/3}-0.006(d/D)^{-7/3}]}{(\sigma_l^2+\sigma_t^2)}\}^{3/5}
\end{equation}

\begin{equation}
\label{Eq_zenith}
    r_0'=r_0\cdot(\cos(z))^{-3/5}
\end{equation}

\begin{figure}[h]
  \begin{minipage}[t]{0.5\linewidth}
  \centering
   \includegraphics[width=45mm,height=60mm]{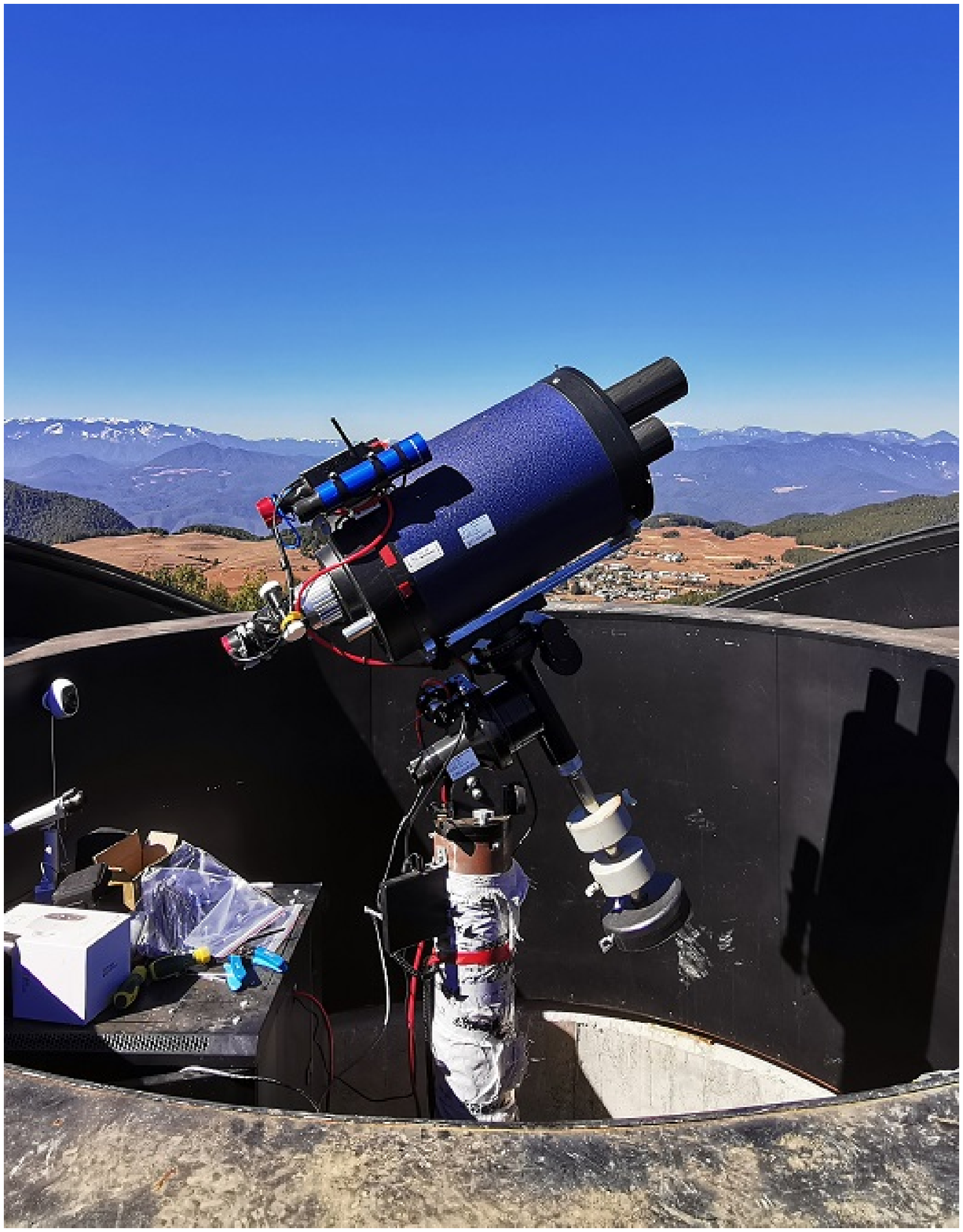}
   \caption{{\small Side of ADIMM System  } }
  \label{fig:adimm01}
  \end{minipage}
  \begin{minipage}[t]{0.5\textwidth}
  \centering
   \includegraphics[width=45mm,height=60mm]{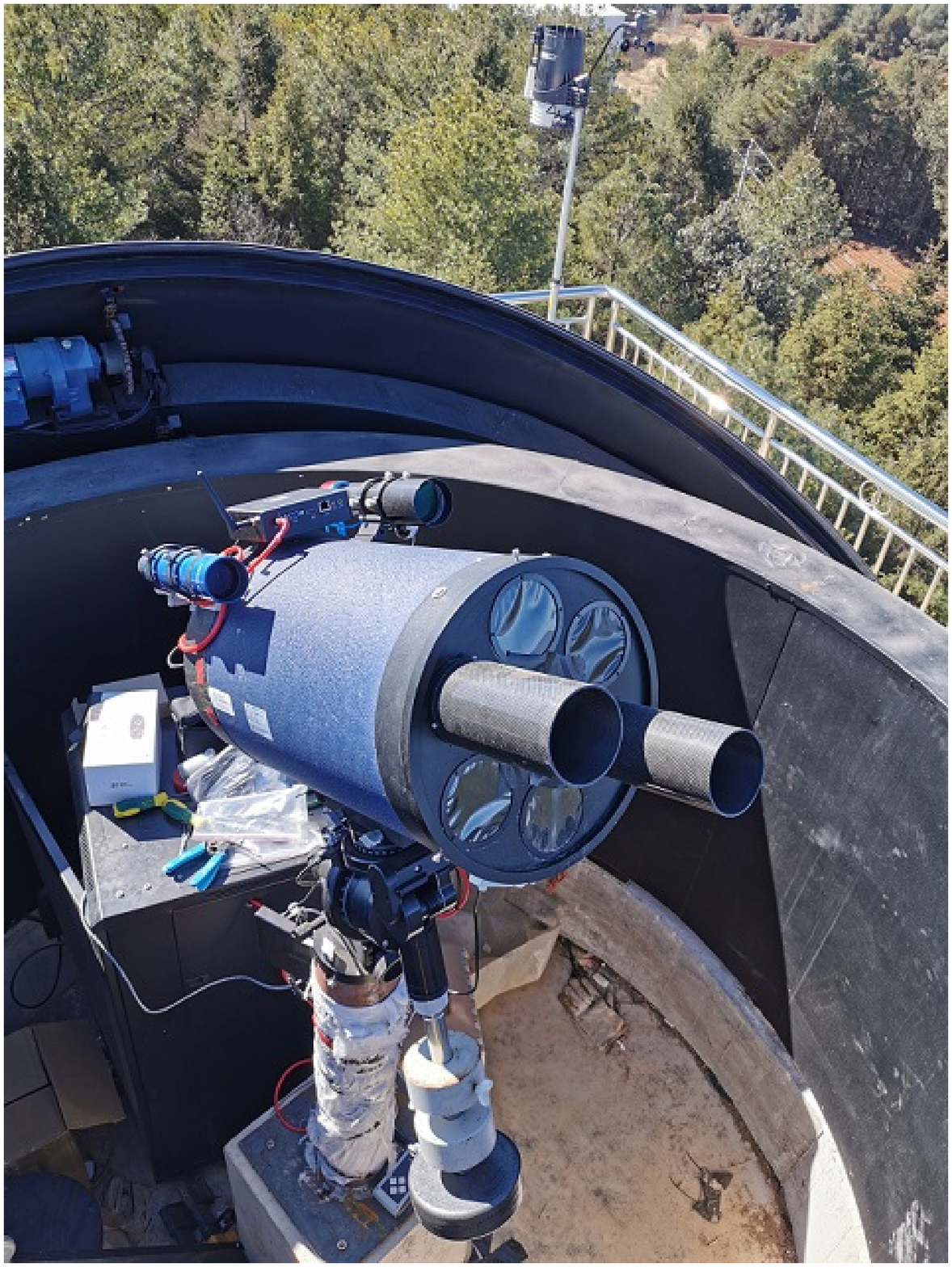}
   \caption{{\small Front of ADIMM System }}
  \label{fig:adimm02}
  \end{minipage}
\end{figure}

We developed a Remote-DIMM (RDIMM) using the traditional formulae and using the wavelength of 550nm, to measure the seeing data of GMG between 2011 and 2018. In early 2019, we upgraded the RDIMM to Autonomous-DIMM (ADIMM) based on the new formulae and used the wavelength of 500nm to get the year-round seeing data at GMG station. ADIMM takes a 30cm telescope and double pupil wedges (Table \ref{table:adimm}) with a double moisture proof carbon fiber cylinder (Fig. \ref{fig:adimm01}, Fig. \ref{fig:adimm02}), for better adaptability to a high humidity environment.

\begin{table}
\caption[]{ADIMM  \label{table:adimm}}
\setlength{\tabcolsep}{1pt}
\small
 \begin{tabular} { p{2.5cm} p{2.5cm} p{3cm} p{3cm} p{3cm} }
  \hline\noalign{\smallskip}
  Telescope & Mount & Dome & Finder Telescope  & Pupils \\
  \hline\noalign{\smallskip}
  Meade 12" F/10 & Losmandy G11 & Clam shell 3m-Dome & Explore 60mm & d=77mm, D=220mm \\
\noalign{\smallskip}\hline
 \end{tabular}
\end{table}

Calibration is critical work for the DIMM system. We select a list of visual binaries (Table \ref{table:stars}) suitable for ADIMM calibration. The selection principle of visual binaries are: (1) they have the similar magnitude, and (2) they have the suitable angular distance. You should recalculate the angular distance at the calibration night. The calibrated parameters of ADIMM are shown in Table \ref{table:adimm02}.

\begin{table}
\bc
\caption[]{Visual Binary Stars \label{table:stars}}
\setlength{\tabcolsep}{1pt}
\small
\begin{tabular} { p{2.2cm} p{3cm} p{3cm} p{3cm} p{3cm} }
\hline\
Object & RA(J2000)/h:m:s & DEC(J2000)/d:m:s & M1,M2/Magnitude & Angular Distance(") \\
\hline\
BDS 71A & 00:14:02.61 & +76:01:37 & 7.1, 7.9 & 76.3 \\
BDS 1094A & 02:12:49.91 & +79:41:29 & 6.5, 7.1 & 55.3 \\
ADS 2984A & 04:07:51.39 & +62:19:48 & 7.0, 7.1 & 17.9 \\
BDS 2867A & 05:37:53.45 & +00:58:07 & 7.2, 7.9 & 80.1 \\
HD 73665 & 08:40:06.4 & +20:00:28 & 6.5, 6.5 & 149.8 \\
HD 77600 & 09:05:45.18 & +50:16:36 & 8.1, 8.3 & 79.2 \\
HD 90125 & 10:24:13.18 & +02:22:04 & 6.4, 6.7 & 212.2 \\
ADS 8434A & 12:08:07.07 & +55:27:50 & 8.0, 8.4 & 22.3 \\
ADS 12815A & 19:41:48.95 & +50:31:30 & 6.3, 6.4 & 39.0 \\
ADS 15670A & 22:08:36.05 & +59:17:22 & 7.2, 7.4 & 21.6 \\
\hline
\end{tabular}
\ec
\end{table}

\begin{table}
\bc
\caption[]{ADIMM  \label{table:adimm02}}
\setlength{\tabcolsep}{1pt}
\small
 \begin{tabular} {p{2cm} p{1.5cm} p{2cm} p{2cm} p{2cm} p{1.5cm} p{1.5cm} p{1.5cm} }
  \hline\noalign{\smallskip}
Camera & Exposure time & Pixel size & System length & Scale & FOV & Frame & Sample \\
  \hline\noalign{\smallskip}
ASI174MM & 5ms & 5.86um & 3164mm & 0.38" pixel$^{-1}$ & 7.7'x12.3' & 197 &2000\\
ASI1600MC & 1s & 3.8um & 232.58mm & 3.37" pixel$^{-1}$ & $3.3^{\circ}$x$4.36^{\circ}$ & 1 & --\\
\noalign{\smallskip}\hline
 \end{tabular}
 \ec
\end{table}

\begin{figure}
   \centering
   \includegraphics[width=0.8\textwidth]{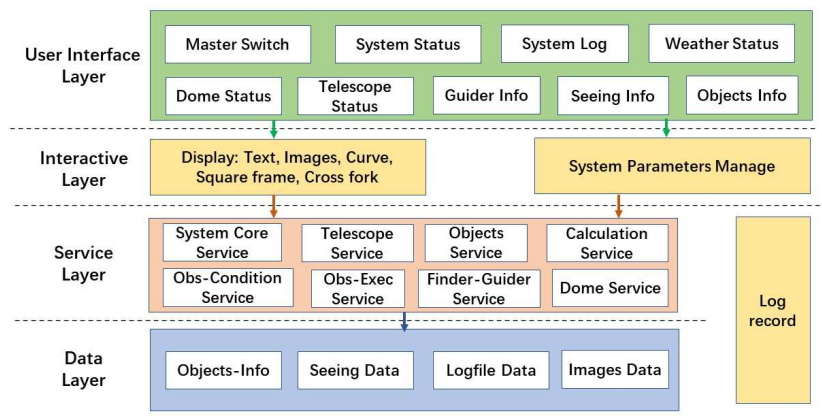}
   \captionsetup{justification=centering}
   \caption{\small Software Architecture of ADIMM System }
   \label{fig:software}
\end{figure}

The ADIMM robotic control software was developed by C++ language, and the software architecture is shown in the Figure \ref{fig:software}. It can automatically open or close the roof by real-time meteorological data and all-sky data,
and also automatically find bright stars for the guiding and measuring the whole atmospheric turbulence. All the seeing data will be automatically uploaded to ASMS Server at each minute during the observable nights.

\subsection{Video Surveillance System}
\label{subsect:VSS}
For the Remote Observatory or Robotic Autonomous Observatory, the video surveillance system for monitoring the dome and telescope is necessary. It is very helpful for troubleshooting. To avoid infrared light pollution, Lijiang Observatory takes the starlight sensitivity camera AXIS1354, for monitoring the telescopes and domes.

\section{Annual Data Analysis}
\label{sect:annual_data_analysis}
\begin{figure}
   \centering
   \includegraphics[width=0.6\textwidth]{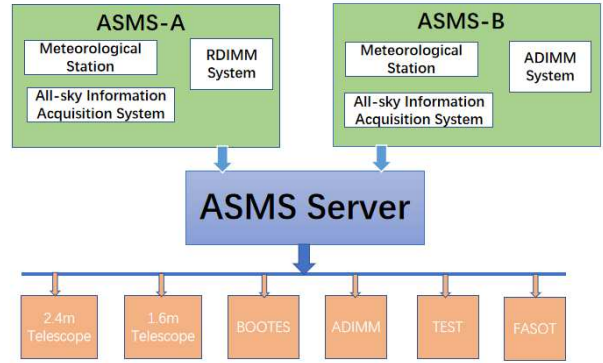}
   \captionsetup{justification=centering}
   \caption{\small Data Flow of ASMS }
   \label{fig:dataflow}
\end{figure}

We use a Linux server for synchronization, storage and online display the data from the ASMS-A and ASMS-B. Meanwhile the server pushes site observation environment data to all the remote or robotic telescopes per minute through the internet. One can browse the current Astroclimatic conditions through the website: \href{http://weather.gmg.org.cn:9000/}{http://weather.gmg.org.cn:9000/}, where all the data are updated every minute. The ASMS data flow is shown in Figure \ref{fig:dataflow}. Through the TCP/IP protocol or FTP protocol, all the telescopes performing at GMG station could get the site information from ASMS-A or ASMS-B in real time.

In 2019, the ASMS-A collected 516 467 sets of weather data, and ASMS-B collected 498 509 sets of weather data (data collection began on 2019 Jan 11). Because the two systems are located at different sites and installed on different heights, the data analysis of the two collections will be carried out simultaneously below.

\subsection{Temperature and Dew Point}
\label{subsect:Temp}

\begin{figure}
  \begin{minipage}[t]{0.5\linewidth}
  \centering
  \setlength{\abovecaptionskip}{0.2cm}
   \includegraphics[width=80mm,height=40mm]{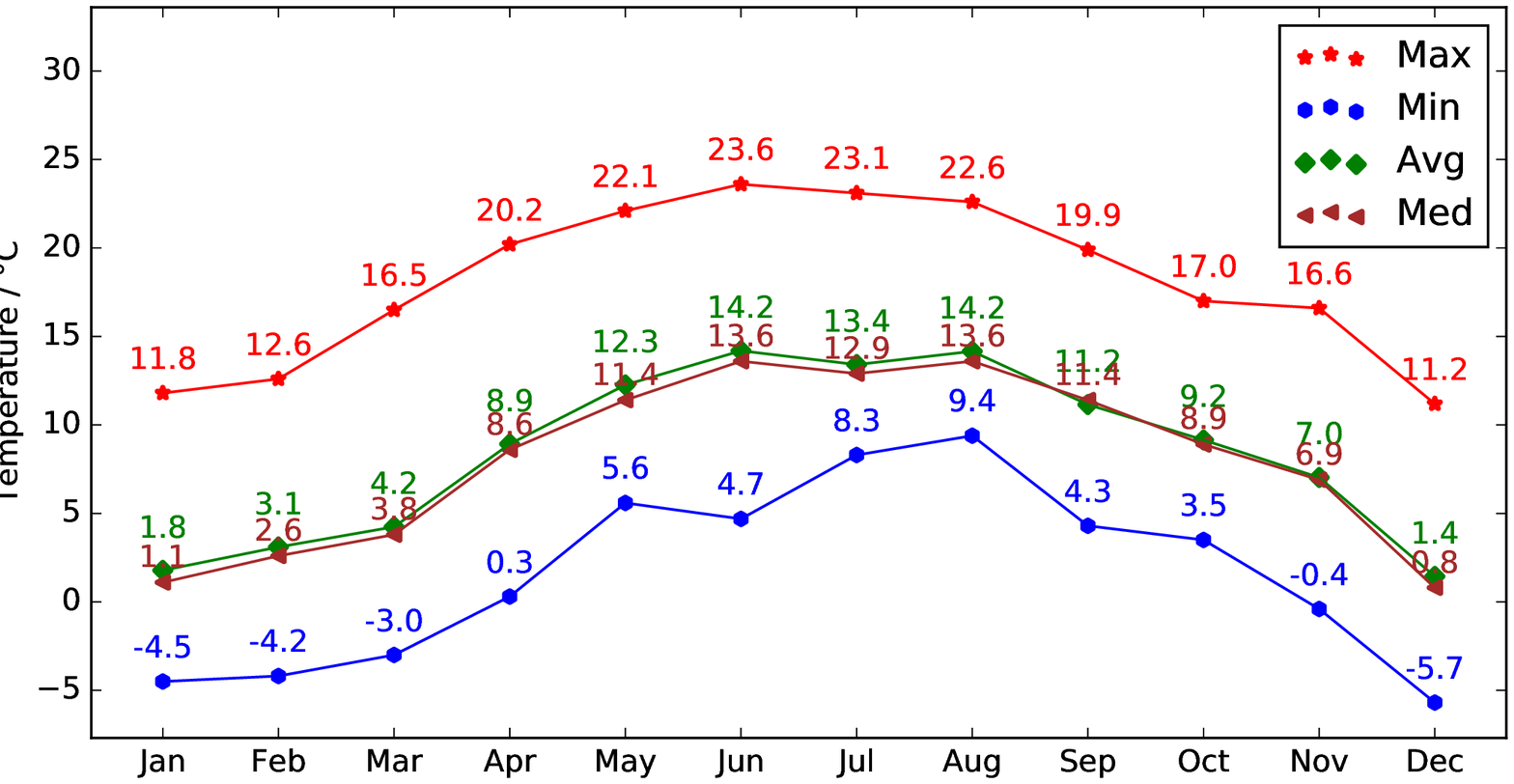}
   \caption{{\small Temperature of ASMS-A in 2019 } }
  \label{fig:temp2m4}
  \end{minipage}
  \begin{minipage}[t]{0.5\textwidth}
  \centering
   \setlength{\abovecaptionskip}{0.2cm}
    \includegraphics[width=80mm,height=40mm]{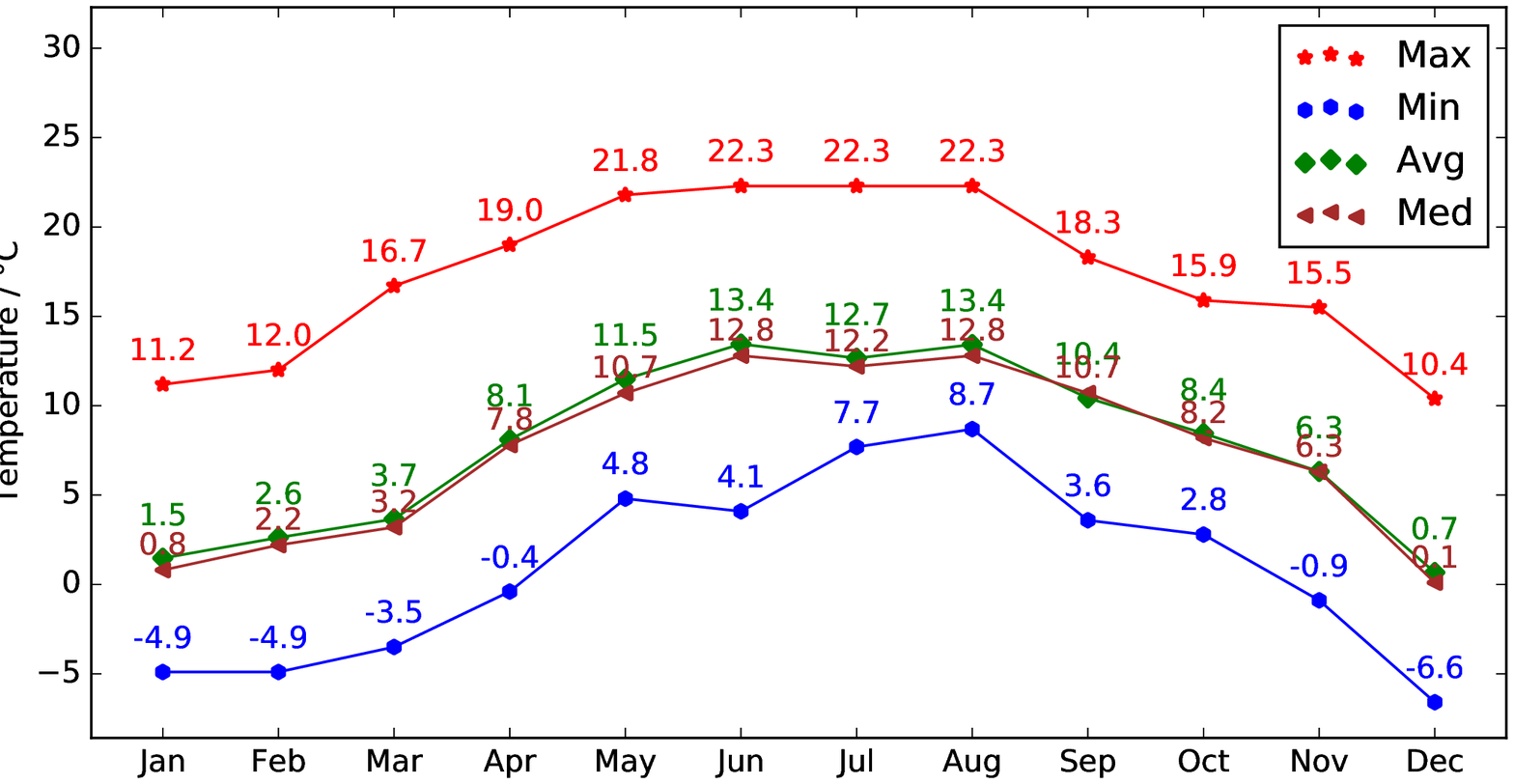}
   \caption{{\small Temperature of ASMS-B in 2019 } }
  \label{fig:temp1m6}
  \end{minipage}
\end{figure}

\begin{figure}
  \begin{minipage}[t]{0.5\linewidth}
  \centering
  \setlength{\abovecaptionskip}{0.2cm}
   \includegraphics[width=80mm,height=40mm]{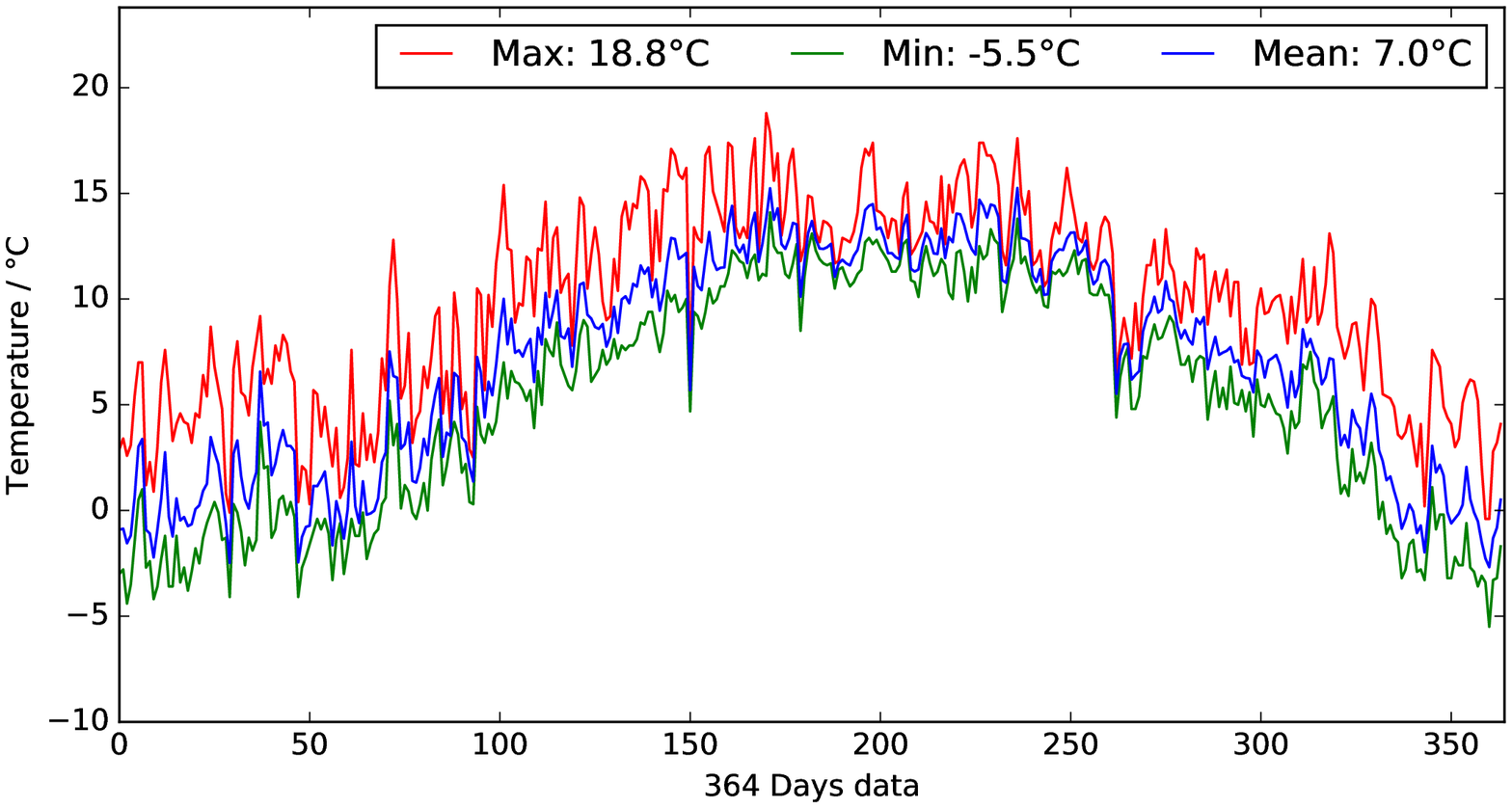}
   \caption{{\small Night Temperature of ASMS-A in 2019}}
  \label{fig:temp2m4_night}
  \end{minipage}
  \begin{minipage}[t]{0.5\textwidth}
  \centering
   \setlength{\abovecaptionskip}{0.2cm}
   \includegraphics[width=80mm,height=40mm]{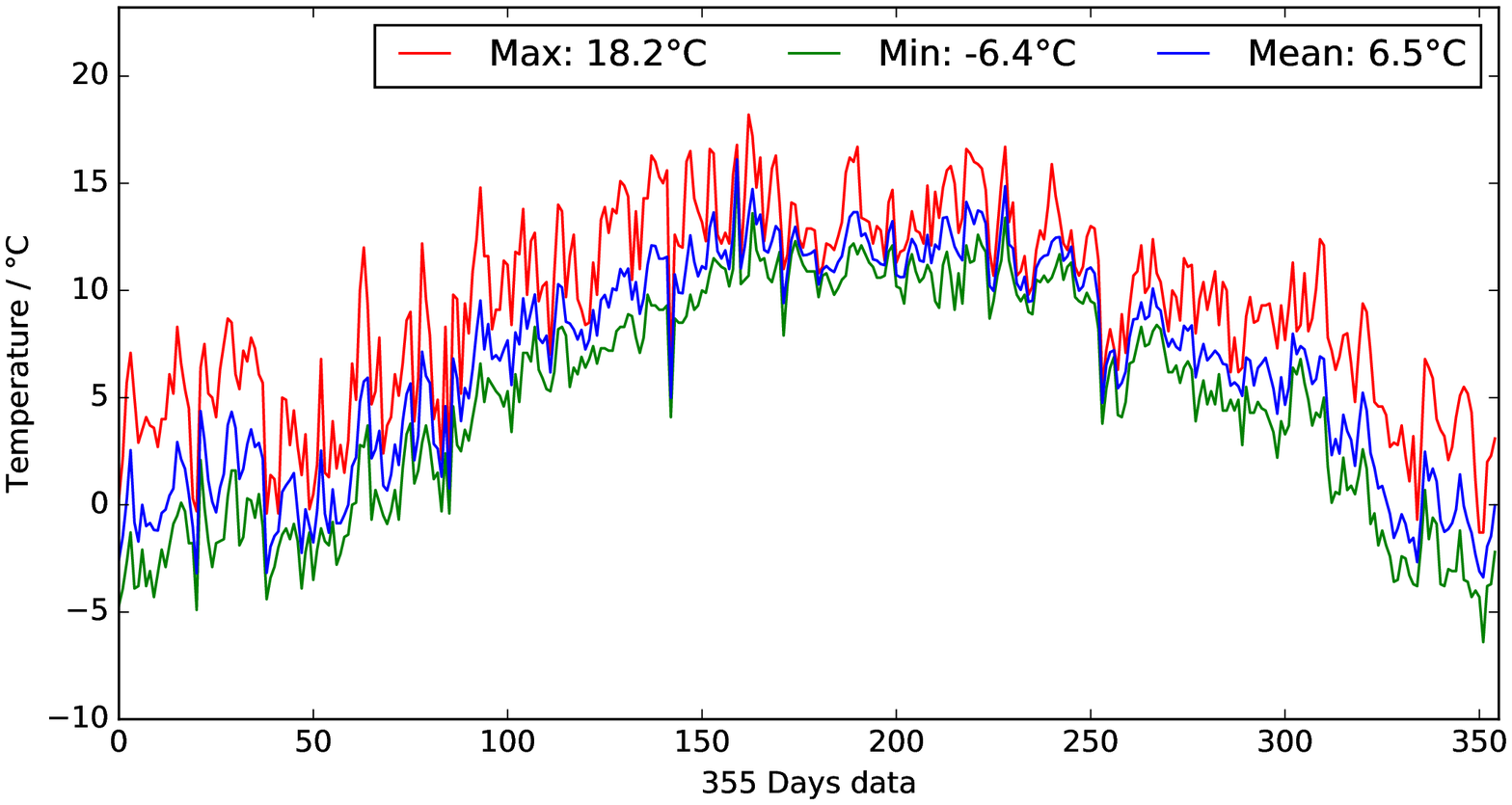}
   \caption{{\small Night Temperature of ASMS-B in 2019}}
  \label{fig:temp1m6_night}
  \end{minipage}
\end{figure}

The average annual temperature is $8.4^{\circ}$C at site A and $7.85^{\circ}$C at site B of Lijiang Observatory. The annual temperature ranges are $-5.7^{\circ}$C $\sim 23.6^{\circ}$C for site A and $-6.6^{\circ}$C $\sim 22.3^{\circ}$C for site B. The range for the whole year is relatively small. The 12-month maximum, minimum, average and median air temperature values are shown on Figure \ref{fig:temp2m4} and Figure \ref{fig:temp1m6}. The hot months are May, June, July and August. The colder months are December, January, February and March.
Lijiang Observatory is mostly a night astronomical observatory, we present the night temperature's maximum, minimum and average on each day by Figure \ref{fig:temp2m4_night} and Figure \ref{fig:temp1m6_night}, where the night means solar altitude is below zero. The night minimum temperature is $-5.5^{\circ}$C for site A and $-6.4^{\circ}$C for site B, which is a little higher than daytime minimum temperature. That because the lowest temperature of the day occurs early in the morning, not the night.


\begin{figure}
  \begin{minipage}[t]{0.5\linewidth}
  \centering
  \setlength{\abovecaptionskip}{0.2cm}
   \includegraphics[width=80mm,height=40mm]{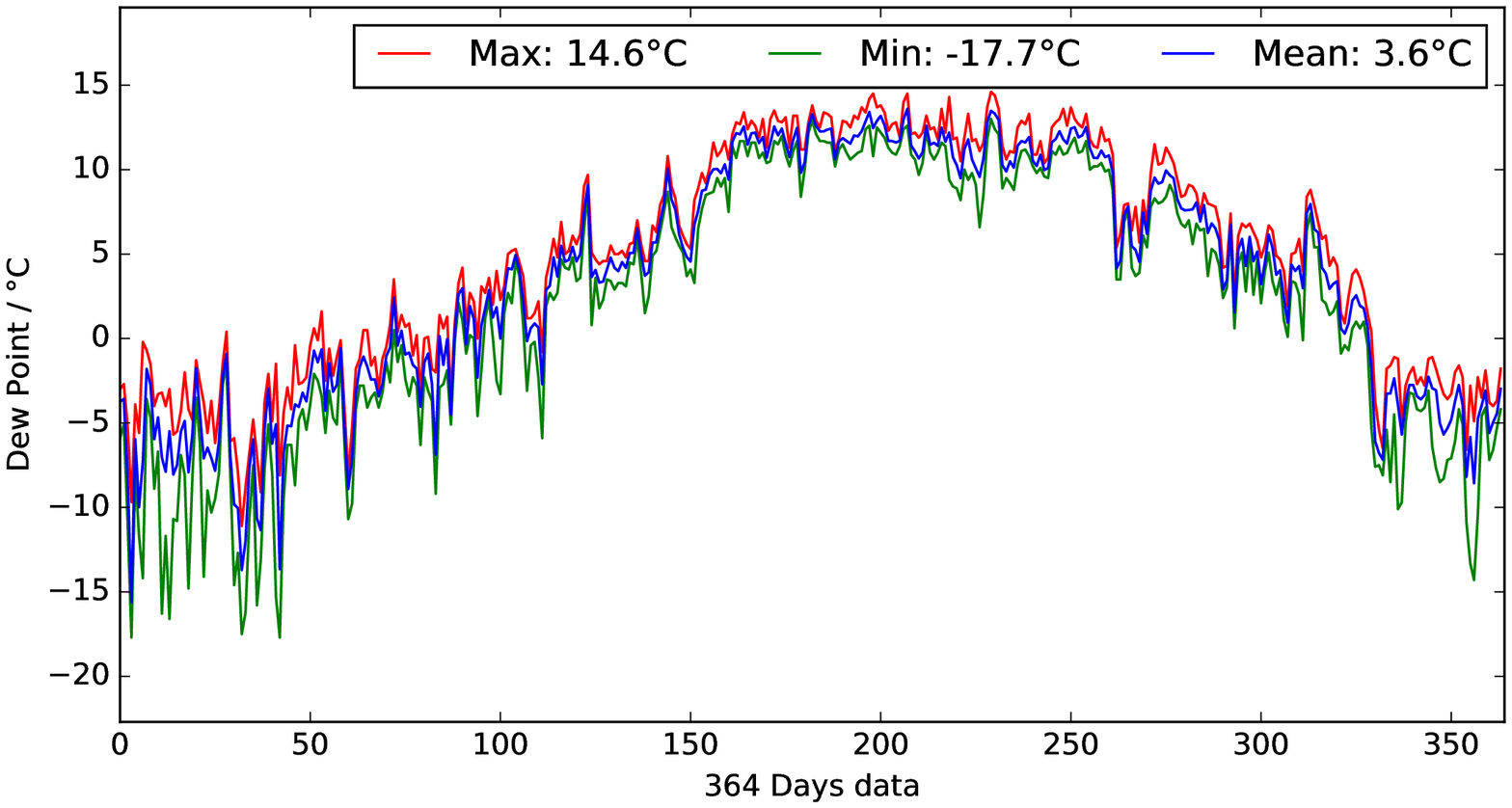}
   \caption{{\small Night Dew Point of ASMS-A in 2019 } }
  \label{fig:dew2m4}
  \end{minipage}
  \begin{minipage}[t]{0.5\textwidth}
  \centering
   \setlength{\abovecaptionskip}{0.2cm}
   \includegraphics[width=80mm,height=40mm]{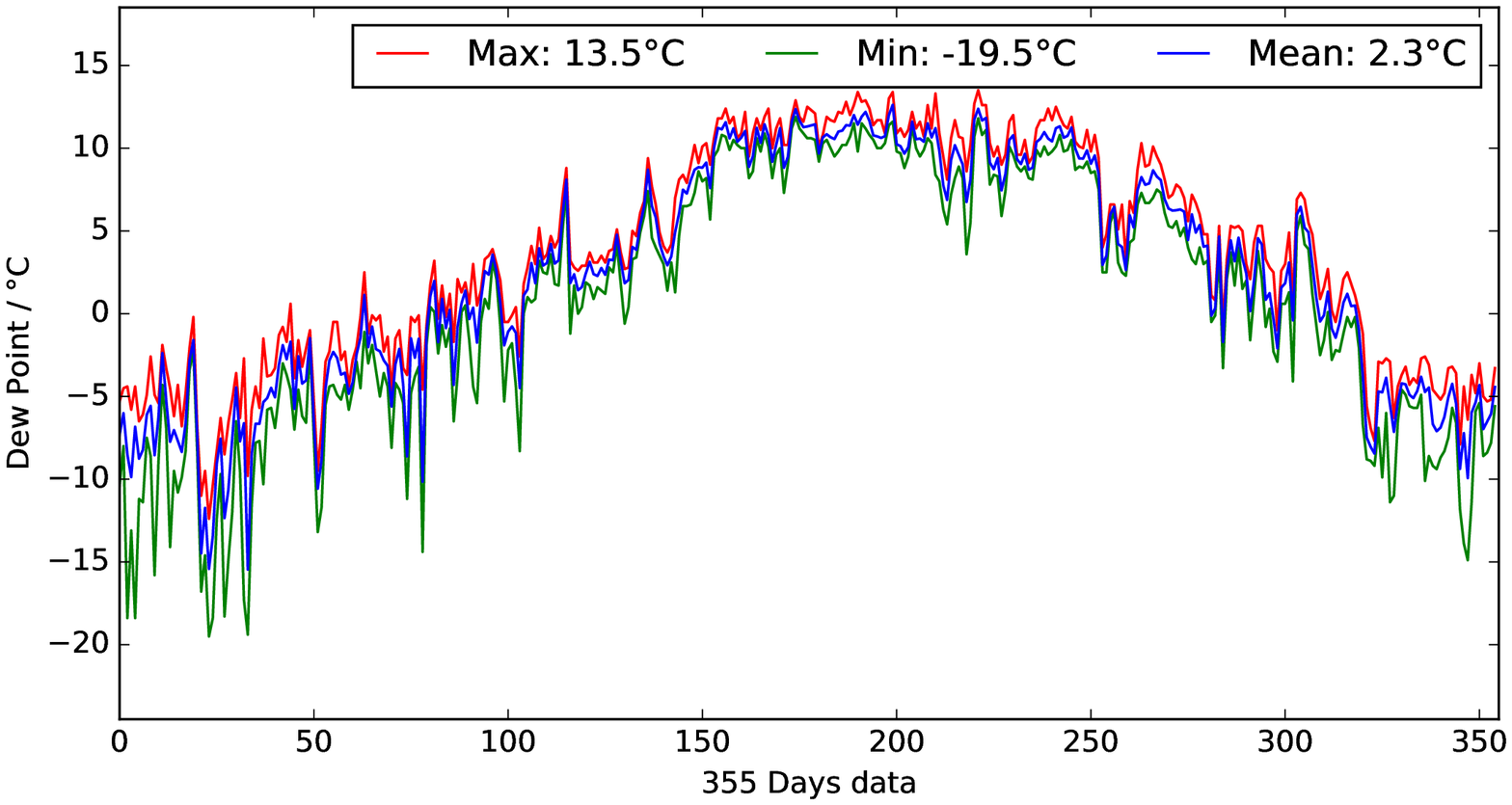}
   \caption{{\small Night Dew Ponit of ASMS-B in 2019}}
  \label{fig:dew1m6}
  \end{minipage}
\end{figure}

Telescope mirrors are easily exposed during the air temperature near dew point, and the dew phenomenon will bring negative effects on the non-protective film of telescope mirrors, such as degradation or shedding. In addition, the water cooler is more and more used for CCD refrigeration, beside the liquid nitrogen. So the dew point parameter is important for a water cooler type selection, to prevent CCD system condensation.
Each night's dew point maximum, minimum, average values are shown in Figure \ref{fig:dew2m4} and Figure \ref{fig:dew1m6}, The maximum is $14.6^{\circ}$C for site A and $13.5^{\circ}$C for site B in summer, and the minimum is $-17.7^{\circ}$C for site A, and $-19.5^{\circ}$C for site B in winter. Therefore the temperature setting values of the water cooler can be adjusted for dew point values in different seasons.

\subsection{Relative Humidity and Precipitation }
\label{subsect:RH}
\begin{figure}
  \begin{minipage}[t]{0.5\linewidth}
  \centering
  \setlength{\abovecaptionskip}{0.2cm}
   \includegraphics[width=80mm,height=40mm]{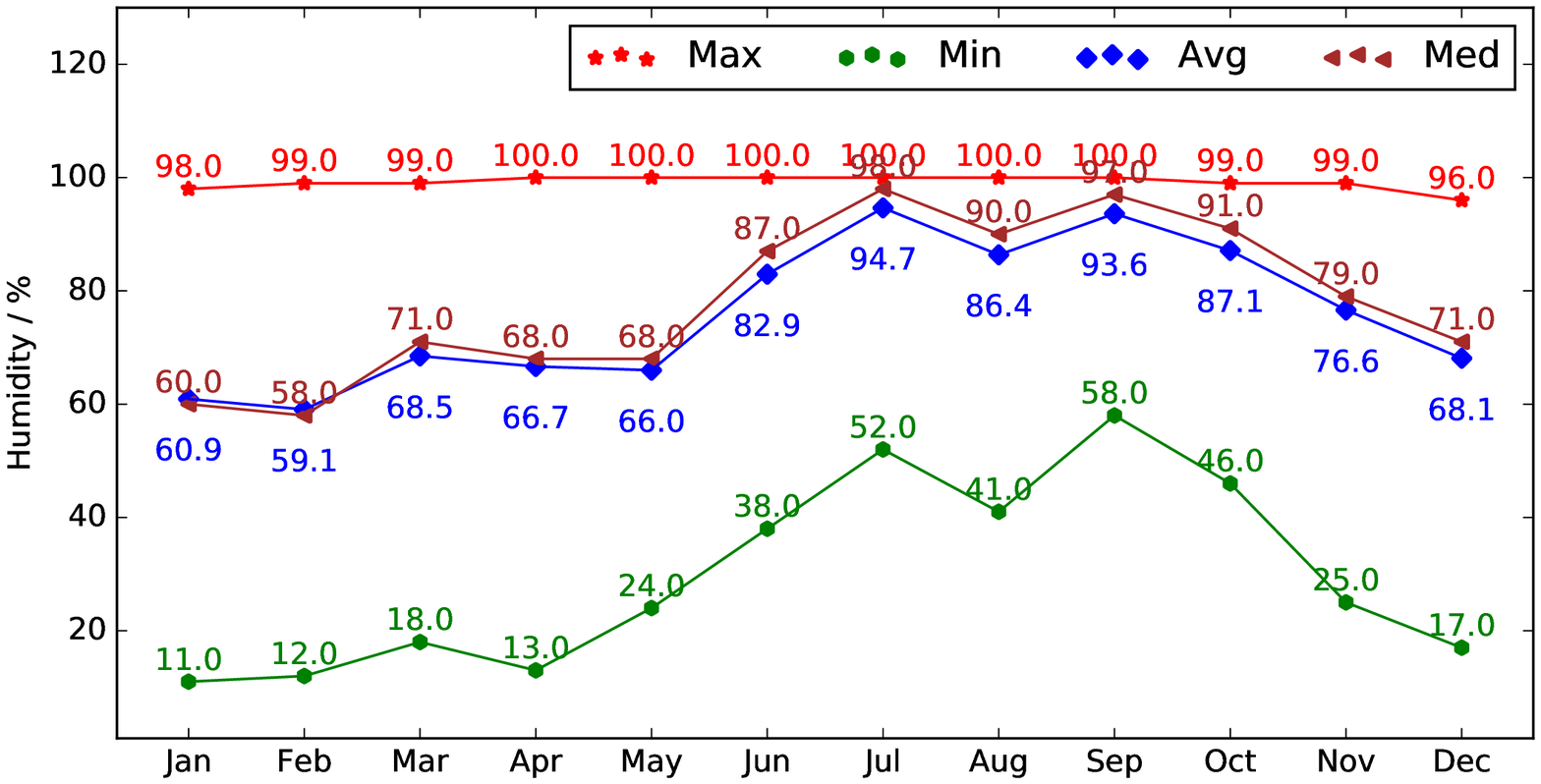}
   \caption{{\small Relative Humidity of ASMS-A in 2019 } }
  \label{fig:rh_2m4}
  \end{minipage}
  \begin{minipage}[t]{0.5\textwidth}
  \centering
  \setlength{\abovecaptionskip}{0.2cm}
   \includegraphics[width=80mm,height=40mm]{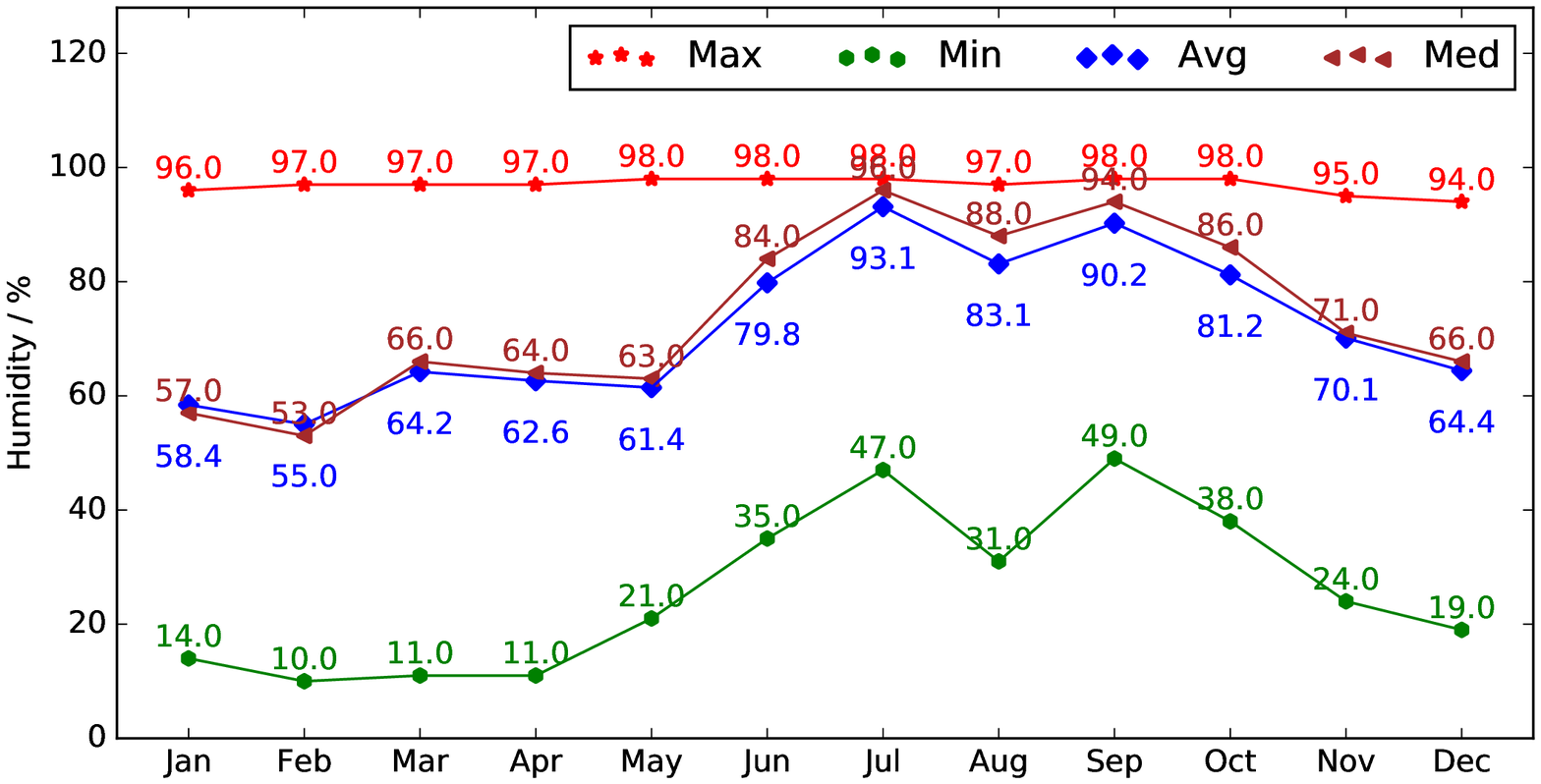}
   \caption{{\small Relative Humidity of ASMS-B in 2019}}
  \label{fig:rh_1m6}
  \end{minipage}
\end{figure}

The 12-month maximum, minimum, average and median relative humidity values are shown in Figure \ref{fig:rh_2m4} and Figure \ref{fig:rh_1m6}. GMG rainy season was postponed in 2019. The typical high humidity time is from June to September. The humidity in October is not relatively high, while the humidity in June is relatively higher. When the humidity is high, the dew point will be very close to the air temperature, and the primary mirror of 2.4m telescope will dew when relative humidity exceeds 95\% outside the dome and 85\% inside the dome. However the full-open dome (such as ADIMM, BOOTES-4, TAT-3, TEST and STT) have to be closed on the condition of relative humidity exceeding 85\%. We have equipped four dehumidifiers inside the 2.4-meter telescope dome for further reducing relative humidity in the dome.

\begin{figure}
  \begin{minipage}[t]{0.5\linewidth}
  \centering
  \setlength{\abovecaptionskip}{0.2cm}
   \includegraphics[width=80mm,height=40mm]{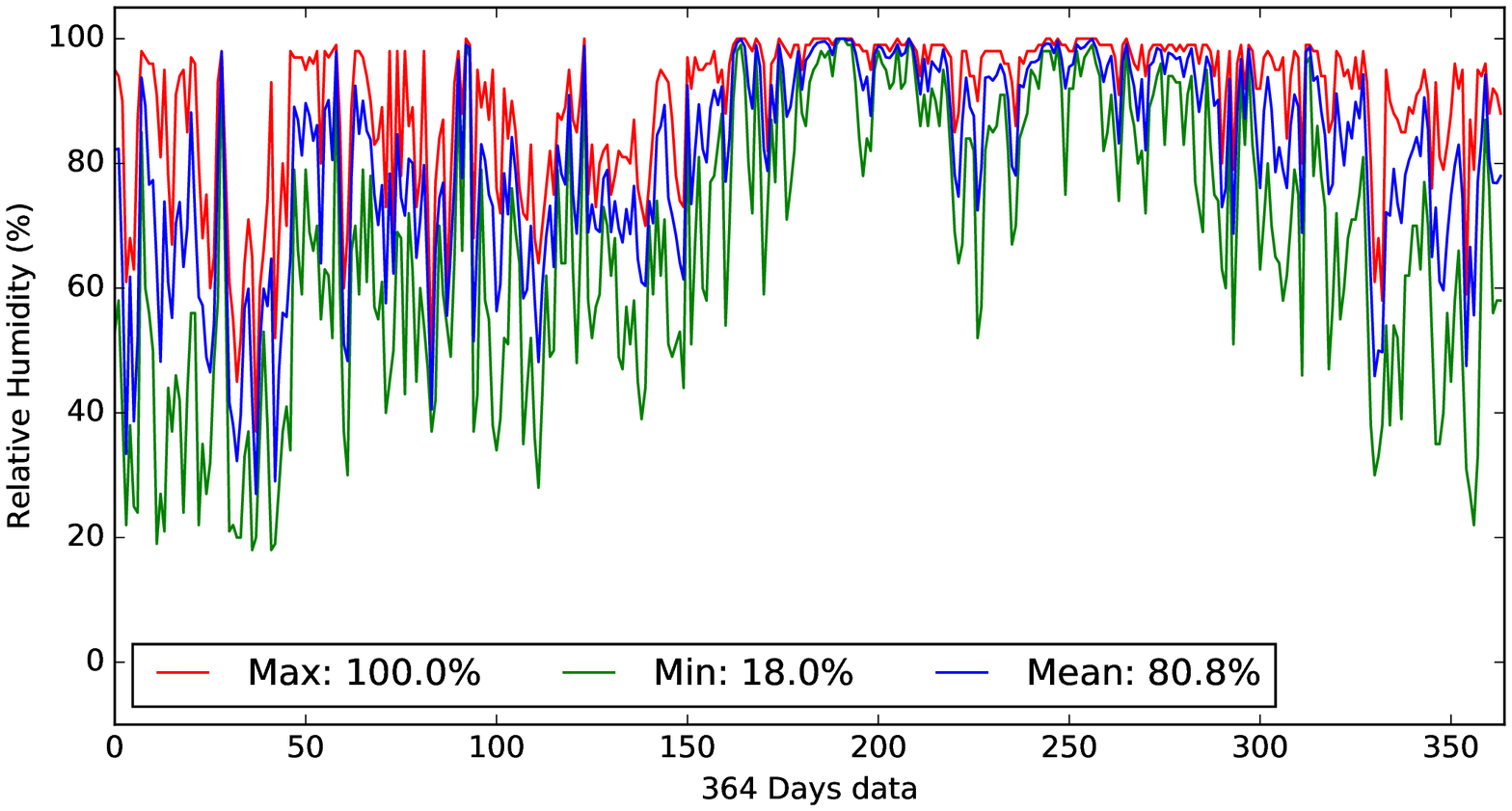}
   \caption{{\small Night Relative Humidity of ASMS-A in 2019 } }
  \label{fig:rh2m4_night}
  \end{minipage}
  \begin{minipage}[t]{0.5\textwidth}
  \centering
  \setlength{\abovecaptionskip}{0.2cm}
   \includegraphics[width=80mm,height=40mm]{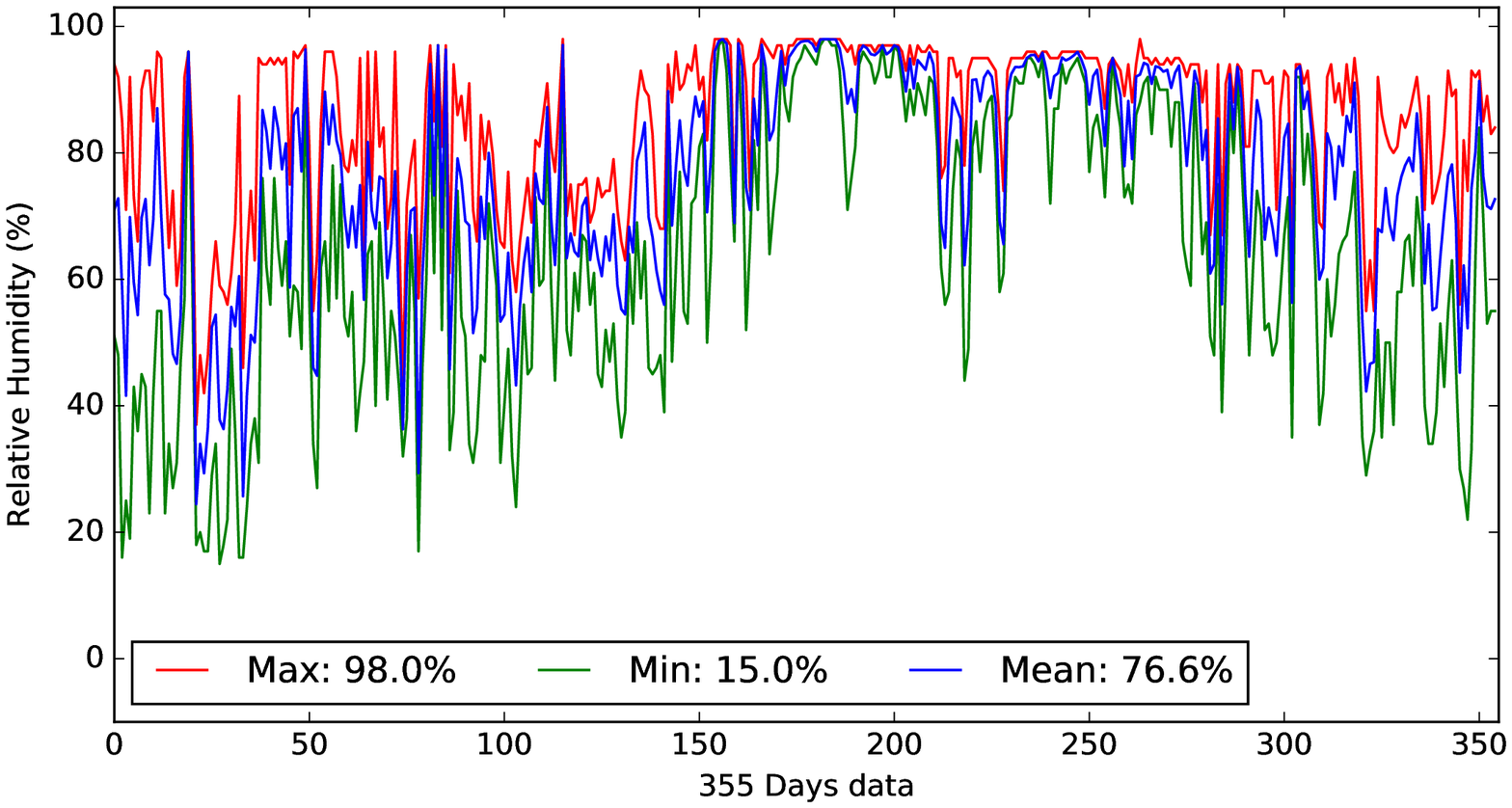}
   \caption{{\small Night Relative Humidity of ASMS-B in 2019}}
  \label{fig:rh1m6_night}
  \end{minipage}
\end{figure}

On each night of the year 2019, the relative humidity maximum, minimum and average values are shown in Figure \ref{fig:rh2m4_night} for site A and Figure \ref{fig:rh1m6_night} for site B. Site B is 8 meters hight than site A (the distance between this two sites is about 300 meters), the average night humidity at site B is 76.6\%, four percent lower than site A. So the higher the telescope pier height, the better, not only for batter seeing, but also better relative humidity. Of course, the telescope pier height need to meet the requirements of the telescope's local oscillator frequency firstly.

\begin{figure}
   \centering
   \includegraphics[width=0.7\textwidth]{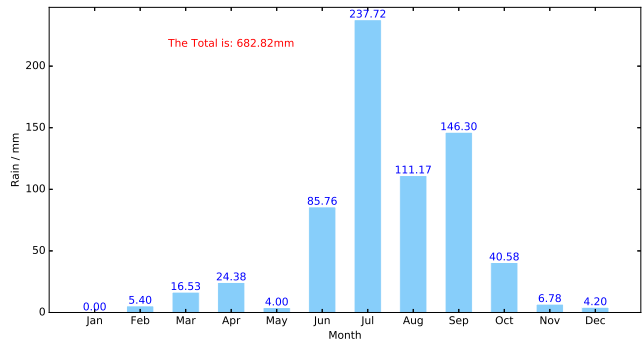}
   \caption{\small Total Rain in 2019 }
   \label{fig:totalrain}
 \end{figure}

Figure \ref{fig:totalrain} shows the annual precipitation distribution, the total rainfall is 682mm in 2019, 85\% rainfall is in June, July, August and September. The four months are known as the raining season in Yunnan province, and the observed nights can be very few.


\subsection{Wind Speed and Wind Direction}
\label{subsect:wind}
\begin{figure}
   \centering
   \setlength{\abovecaptionskip}{0.2cm}
   \includegraphics[width=0.8\textwidth]{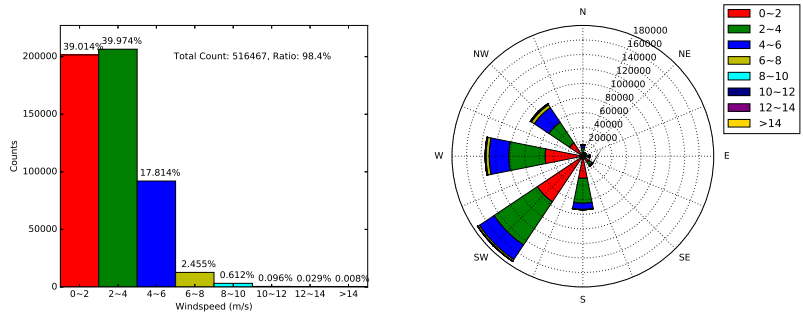}
   \caption{\small Wind Speed and Wind Direction of ASMS-A in 2019 }
   \label{fig:windrose01}
 \end{figure}

\begin{figure}
   \centering
   \setlength{\abovecaptionskip}{0.2cm}
   \includegraphics[width=0.8\textwidth]{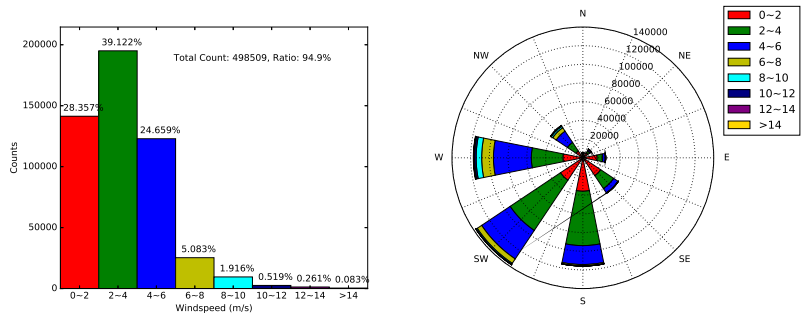}
   \caption{\small Wind Speed and Wind Direction of ASMS-B in 2019 }
   \label{fig:windrose02}
 \end{figure}

Wind speed and wind direction for site A and B are shown on Figure \ref{fig:windrose01} and Figure \ref{fig:windrose02}. 96.8\% of the wind speed are less than 6m s$^{-1}$ at site A and 92.1\% at site B, The 2.4-meter Telescope can be operated under the wide speed of 15m s$^{-1}$ with good image quality, so the wind condition is nearly almost suitable for LJT, and more than 90\% time suitable for smaller telescopes at wind speed below 10m s$^{-1}$ . The yearly maximum wind gust is 22m s$^{-1}$ recorded on February, and the high speed wind is usually in the evening within a day. The yearly average wind is 3.2m s$^{-1}$. Southwest wind is the main direction in the GMG station, and the southwest monsoon climate also impacts the rainfall distribution and observable nights in this astronomical site.


\subsection{Air Pressure}
\label{subsect:pressure}
Air pressure, like the water pressure, is related to the weight of the medium(air or water) over a given height or depth. The air pressure decreases as the altitude increase. The pressure at a given altitude (up to 50km) can be calculated using the Earth Atmosphere Model\footnote{https://www.mide.com/air-pressure-at-altitude-calculator}. Equation \textasciitilde \ref{Eq_pressure01} is applied to Troposphere and Upper Stratosphere, and equation \textasciitilde \ref{Eq_pressure02} is applied to Lower Stratosphere. We simplified the model as shown in Equation \ref{Eq_pressure03}, based on the standard parameters on sea level: $h_b=0$ m, $T_b=288.15$ K$=15^{\circ}$C, and $P_b=101 330$ Pa.

\begin{equation}
\label{Eq_pressure01}
   P=P_b\cdot [1+\frac{L_b}{T_b}\cdot (h-h_b)]^\frac{-g_0\cdot M}{R\cdot L_b}
\end{equation}

\begin{equation}
\label{Eq_pressure02}
  P=P_b\cdot exp[\frac{-g_0\cdot M \cdot(h-h_b)}{R \cdot T_b}]
\end{equation}

where $P_b$ is static pressure at the bottom of atmospheric layer[Pa], $T_b$ is the standard temperature at the bottom of atmospheric layer[K],$L_b$ is the standard temperature lapse rate $=-0.0065 K\cdot m^{-1}$ at the Troposphere and $-0.00299 K\cdot m^{-1}$ at the Lower Stratosphere, $h_b$ is the height at the bottom of atmospheric layer[m],h is the height about sea level[m],$g_0$ is the gravitational acceleration constant$=9.80665 m\cdot s^{-2}$,R is the universal gas constant$=8.31432\frac{\rm N\cdot m}{\rm mol\cdot K}$, M is the molar mass of Earth's air$\rm =0.0289644 Kg\cdot mol^{-1}$.

\begin{equation}
\label{Eq_pressure03}
\begin{aligned}
P
&=101330*[\frac{T+273.15}{288.15}]^{5.25588},\quad T=15.0-0.00649\cdot h,\quad \mbox{\rm h} \mbox{$\in$[0km,11km]} \\
&=22633.2*exp(1.73457-0.0001576h),\quad  T=-56.5, \quad \mbox{\rm h}  \mbox{$\in$[11km,25km]}\\
&=2488.75*[\frac{T+273.15}{216.65}]^{-11.4258},\quad T=-(131.25-0.00299\cdot h), \quad \mbox{\rm h} \mbox{$\in$[25km,50km]}\\
\end{aligned}
\end{equation}

\begin{figure}
   \centering
   \includegraphics[width=140mm,height=80mm]{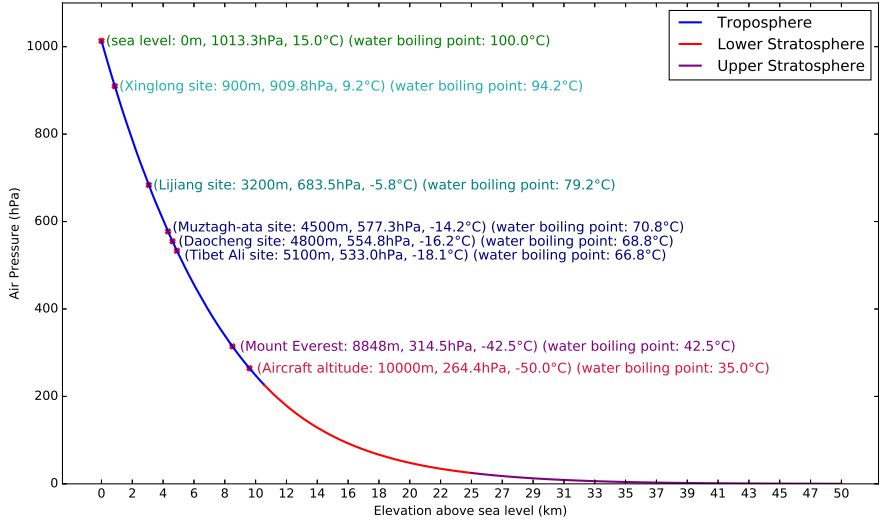}
   \caption{\small Pressure at Troposphere and Stratosphere }
   \label{fig:pressure_whole}
 \end{figure}

Figure \ref{fig:pressure_whole} shows the pressures and the boiling points of water at several optical astronomical sites of China and other special locations, Muztagh-ata(\citealt{site_mstg}), Daocheng(\citealt{site_daocheng}) and Tibet Ali sites(\citealt{site_ali}) are three candidate locations for the 12-meter Large Optical/infrared Telescope(LOT) under the National Fourteenth Five-year Plan in China.

\begin{figure}
  \begin{minipage}[t]{0.5\linewidth}
  \centering
   \includegraphics[width=80mm,height=40mm]{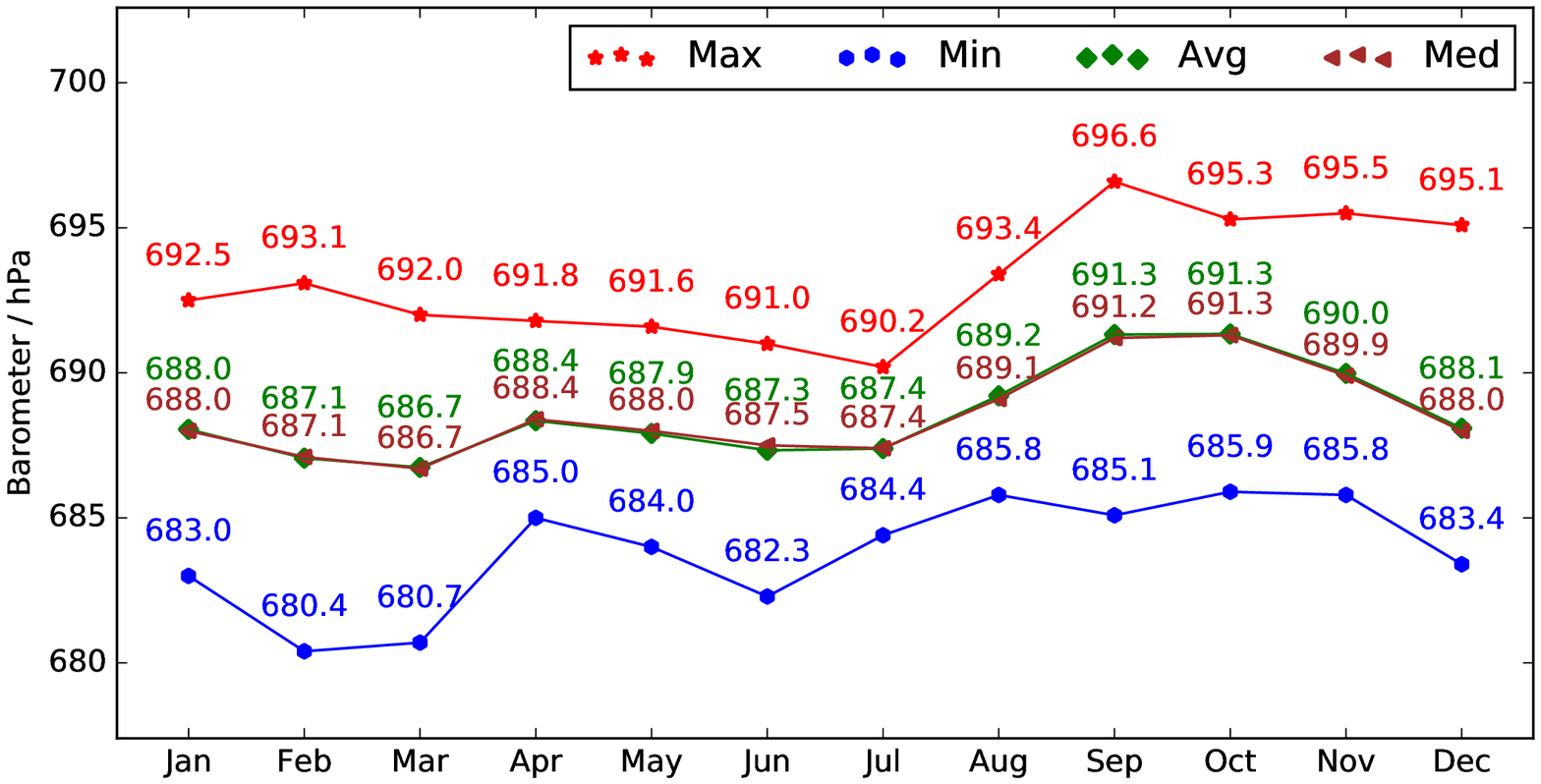}
   \caption{{\small Air Pressure of ASMS-A in 2019 } }
  \label{fig:pressure_a}
  \end{minipage}
  \begin{minipage}[t]{0.5\textwidth}
  \centering
   \includegraphics[width=80mm,height=40mm]{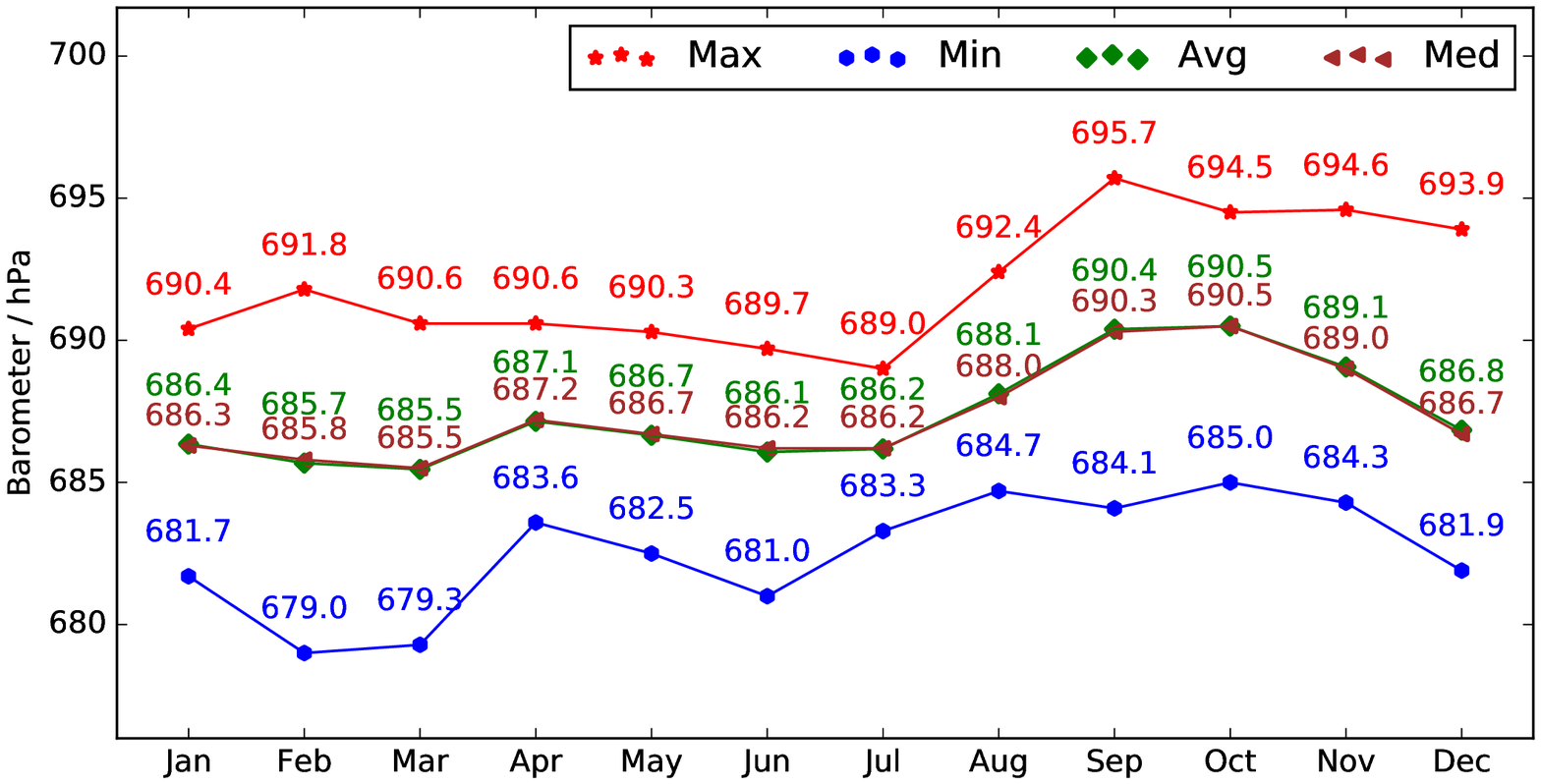}
   \caption{{\small Air Pressure of ASMS-B in 2019}}
  \label{fig:pressure_b}
  \end{minipage}
\end{figure}

Lijiang Observatory's pressure changes throughout the whole year are shown at Figure \ref{fig:pressure_a} from ASMS-A and Figure \ref{fig:pressure_b} from ASMS-B, the minimum pressure is 679hPa, the maximum pressure is 696.6hPa, the average pressure is 687.38hPa. Higher air pressure is in autumn and winter, and lower air pressure is in spring and summer. Within a day, higher air pressure at the midnight, and lower air pressure at evening while the wind gust also happened.

\subsection{UV Index and Solar Radiation}
\label{subsect:UV}
There are two solar coronagraph telescopes at the GMG station (see Table \ref{table:telescopes}), so the UV Index (UVI,the level of UV radiation) and Solar Radiation are helpful observable condition parameters in Daytime. We select 12 sunny days with approximate equal intervals in the whole year, for getting the site UV and solar radiation condition .

\begin{figure}
  \begin{minipage}[t]{0.5\linewidth}
  \centering
   \includegraphics[width=80mm,height=60mm]{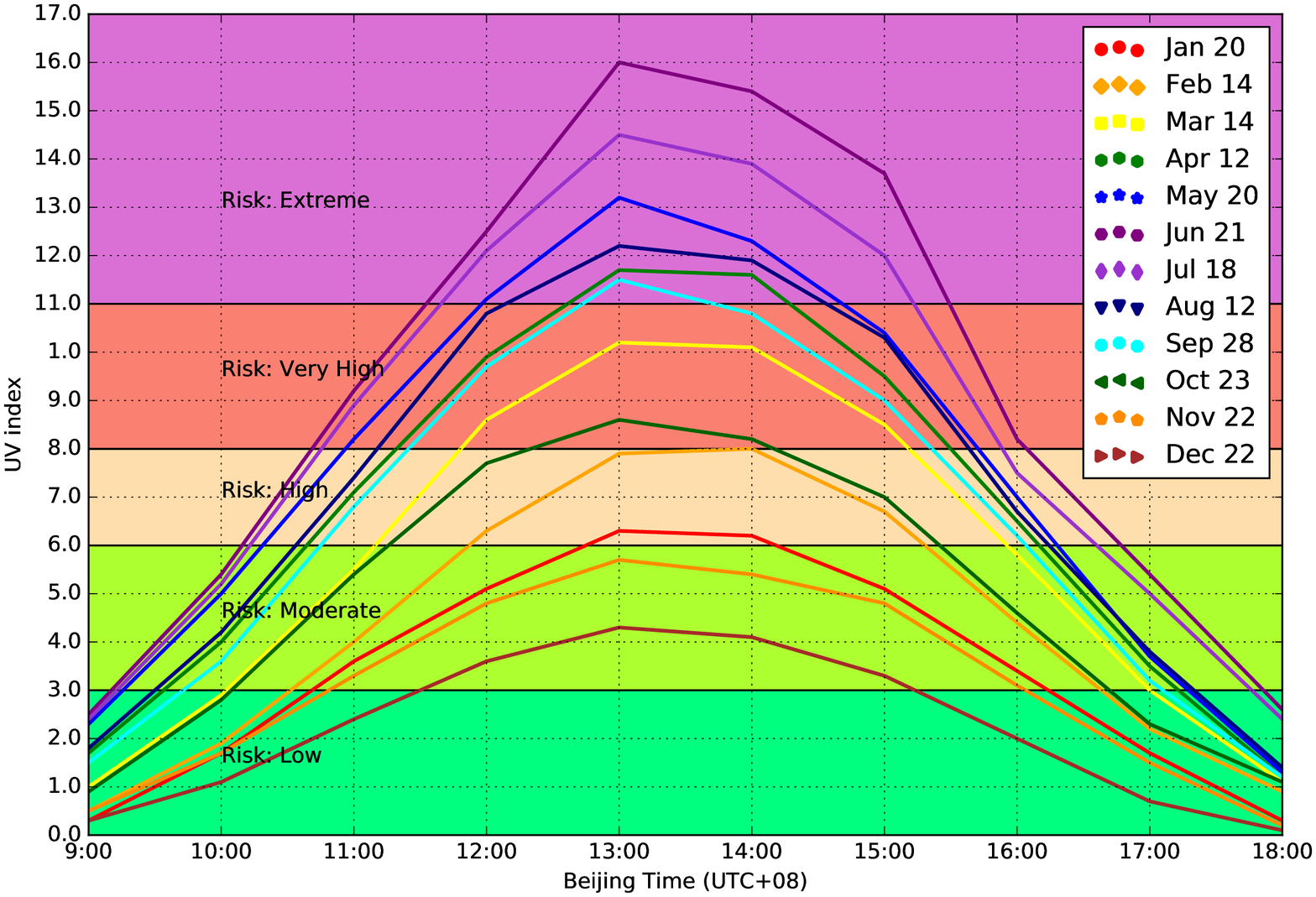}
   \caption{\small UV Index in 2019 }
  \label{fig:UV}
  \end{minipage}
  \begin{minipage}[t]{0.5\textwidth}
  \centering
   \includegraphics[width=80mm,height=60mm]{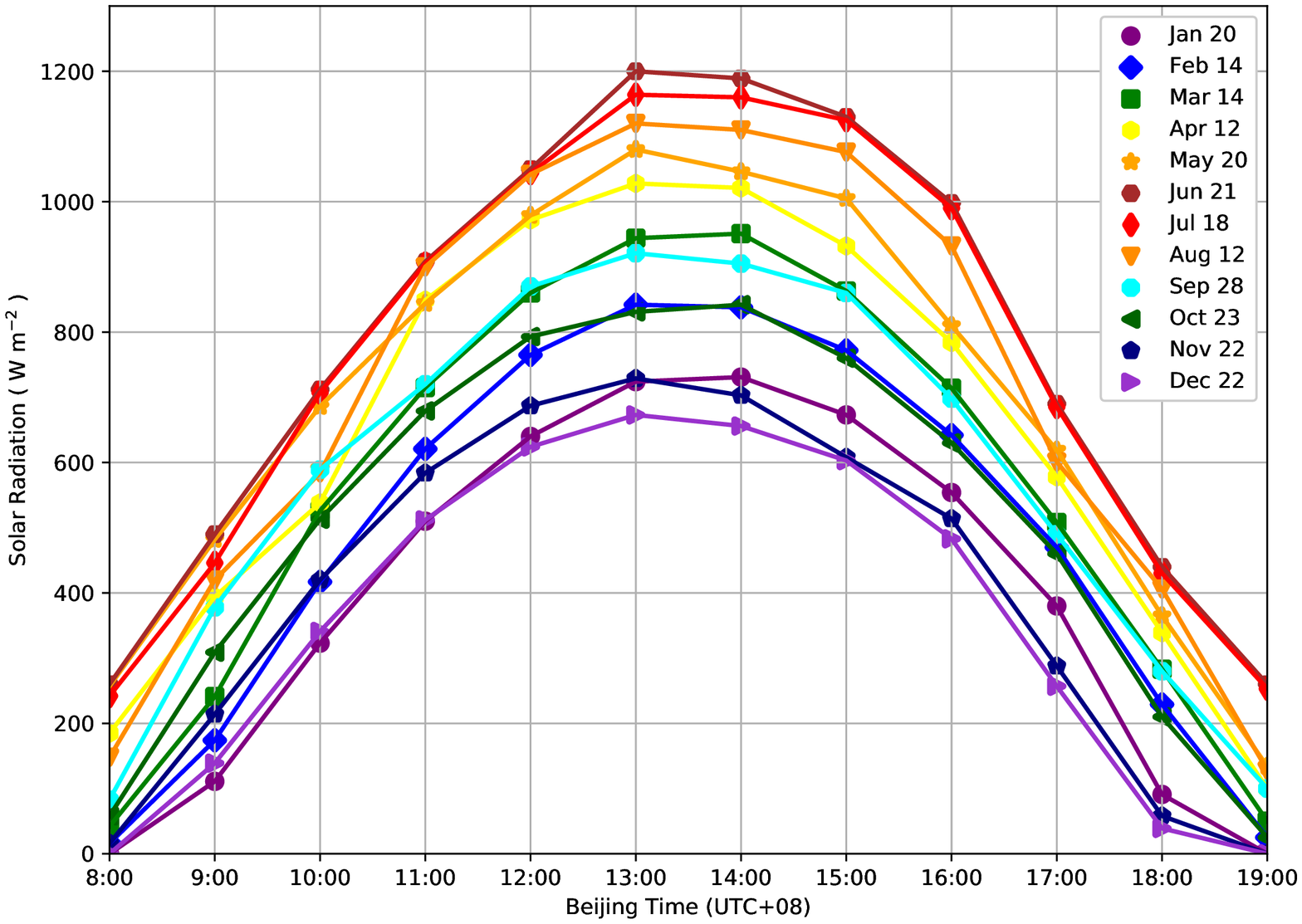}
   \caption{\small Solar Radiation in 2019}
  \label{fig:Solar}
  \end{minipage}
\end{figure}

Figure \ref{fig:UV} shows the UV Index changes in 12 days across the year, and the maximum change from $4.4$ on December to $16$ on June. The UVI has five risk levels established by the World Health Organization, such as Low($0\sim2$), Moderate($3\sim5$), High($6\sim7$), Very High($8\sim10$) and Extreme(11+), and they are marked by different colors as the background shown in Figure \ref{fig:UV}. Only the Low UV level(green area) is no protection needed, we suggest the observers to use the sun protection (generously apply broad-spectrum SPF-15 or higher sunscreen and sunglasses) during the daytime in Lijiang Observatory.

Sunlight is the most influence factor for local weather condition on the earth. For example,the sun's 11-year cycle directly affects the warmth of the earth. Actually only 46\% of incoming sunlight reaches the earth ground
and the reminder is absorbed or scattered by the atmosphere and cloud,
so the solar radiation monitoring can used to calculate the solar disk cloud cover in real time. Figure \ref{fig:Solar} shows solar radiation variation in 12 days across the year. The maximum solar radiation is from 677W m$^{-2}$ in December to 1200W m$^{-2}$ in June, and which led to the temperature changes in the whole year (see Fig.\ref{fig:temp2m4}).

\subsection{Sky Brightness and Atmospheric Extinction}
\label{subsect:darkness}
\begin{figure}
   \centering
   \includegraphics[width=0.8\textwidth]{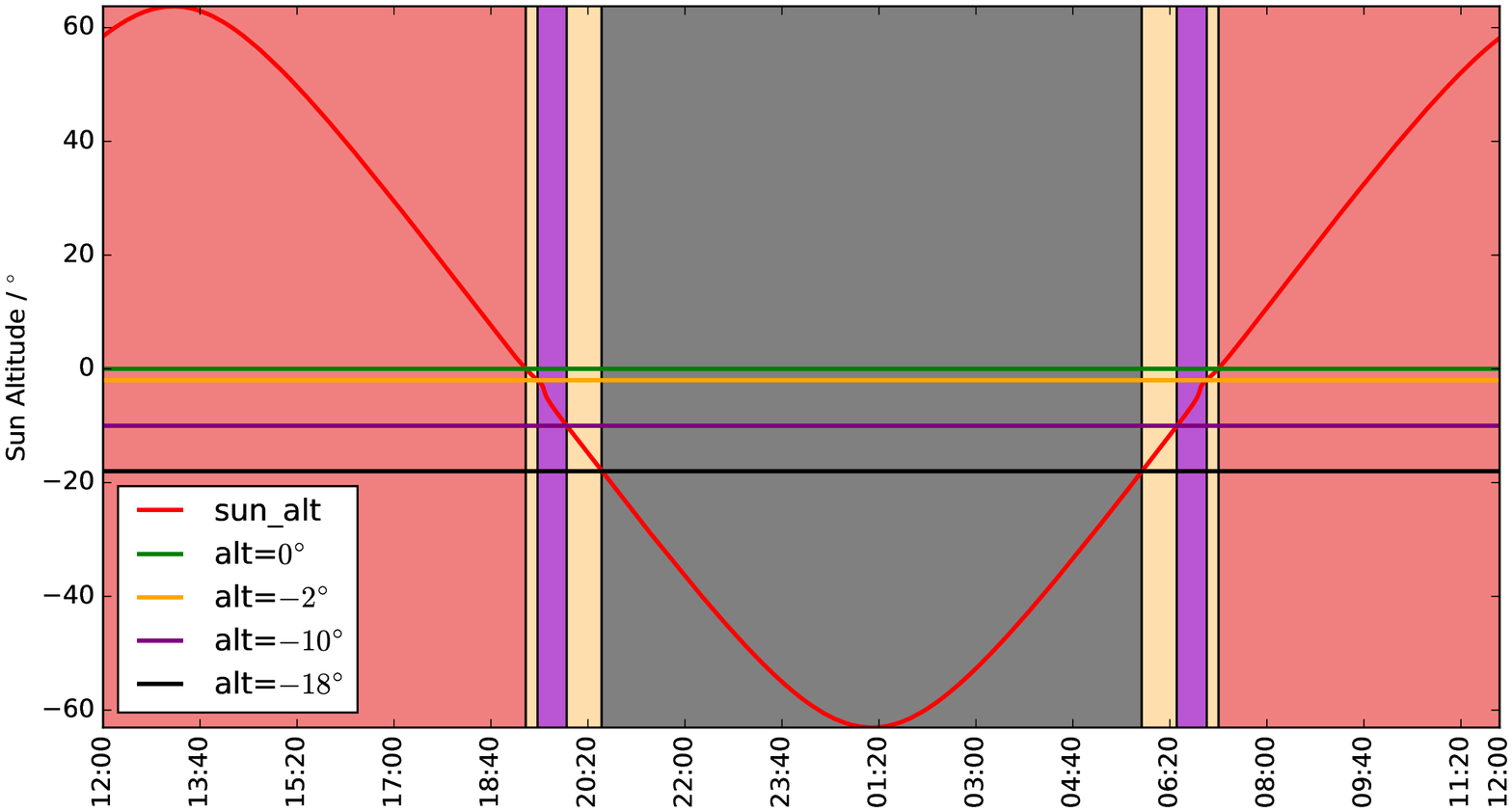}
   \captionsetup{justification=centering}
   \caption{\small GMG Time Blocks on 2019-09-22 \\ (red area: Daytime, orange area: Astronomical twilight time, purple area: Flat time, black area: Night time) }
   \label{fig:timeblocks}
 \end{figure}

There are four time blocks for one site in the Earth rotation cycle: Daytime, Twilight, Night and Twilight. The twilight is the period between dawn and sunset or between sunset and dusk. It can also be separated in civil, nautical and astronomical sections at three sun altitude: $-6{^\circ}$, $-12{^\circ}$ and $-18{^\circ}$. The astronomical night begins at the end of astronomical twilight, and is finished at start of next astronomical twilight.
Sky flat images are required data for photometric observations. It can only be observed between Sun heights between -2 and -10 at the GMG station, the observable time is just 17 minutes in twilight section. We call it Flat time (see Fig. \ref{fig:timeblocks}).

In Table \ref{table:darkness}, we compare the maximum and minimum sky brightness on Johnson $B$, Johnson $V$ broadband and SQM data with several world famous optical observatories. It is shown that moonless sky brightness is directly related to the sites altitude. GMG's darkness condition is similar to San Pedro M\'{a}rtia(SPM) in M\'{e}xico and La Palma in Spain.

\begin{table}
\bc
\caption[]{ Comparison of Zenith Sky brightness at different Sites \label{table:darkness}}
\setlength{\tabcolsep}{1pt}
\begin{tabular} {p{1.5cm} p{1.5cm} p{2cm} p{1.3cm} p{1.3cm} p{1.3cm} p{1.3cm} p{1.3cm} p{1cm}}
\hline
Site & Year & Elevation (m) & JB$_{max}$ & JV$_{max}$ & JB$_{min}$ & JV$_{min}$ & SQM & Ref \\
\hline
GMG & 1996 & 3200 & 22.79 & 21.77 & 21.92 & 21.10 & - & 1 \\
La Silla & 1996 & 2400 & 22.97 & 22.02 & 22.2 & 20.85 & - & 2 \\
La Palma & 1998 & 2400 & 22.70 & 21.90 & - & - & - & 3 \\
Hawaii & 1997 & 4200 & 22.87 & 21.91 & 22.19 & 21.29 & - & 4 \\
SPM & 2015 & 2800 & 22.70 & 21.64 & 22.39 & 21.45 & 21.88 & 5 \\
XingLong & 2012 & 900 & - & - & - & - & 21.1 & 6 \\
GMG & 2019 & 3200 & - & - & - & - & 21.89 & 7 \\
\hline
\end{tabular}
\tablecomments{\textwidth}{Note: The unit is $Mag/arcsec^{2}$, and the SQM value have been subtracted the offset 0.08. Ref: 1. \citealt{GMG1999}, 2. \citealt{Lasilla1996}, 3. \citealt{Lapalma1998}, 4. \citealt{Hawaii1997}, 5. \citealt{SPM2015}, 6. \citealt{xinglong+2015}, 7. this work. }
\ec
\end{table}

\begin{figure}
  \begin{minipage}[t]{0.5\linewidth}
  \centering
   \includegraphics[width=70mm,height=40mm]{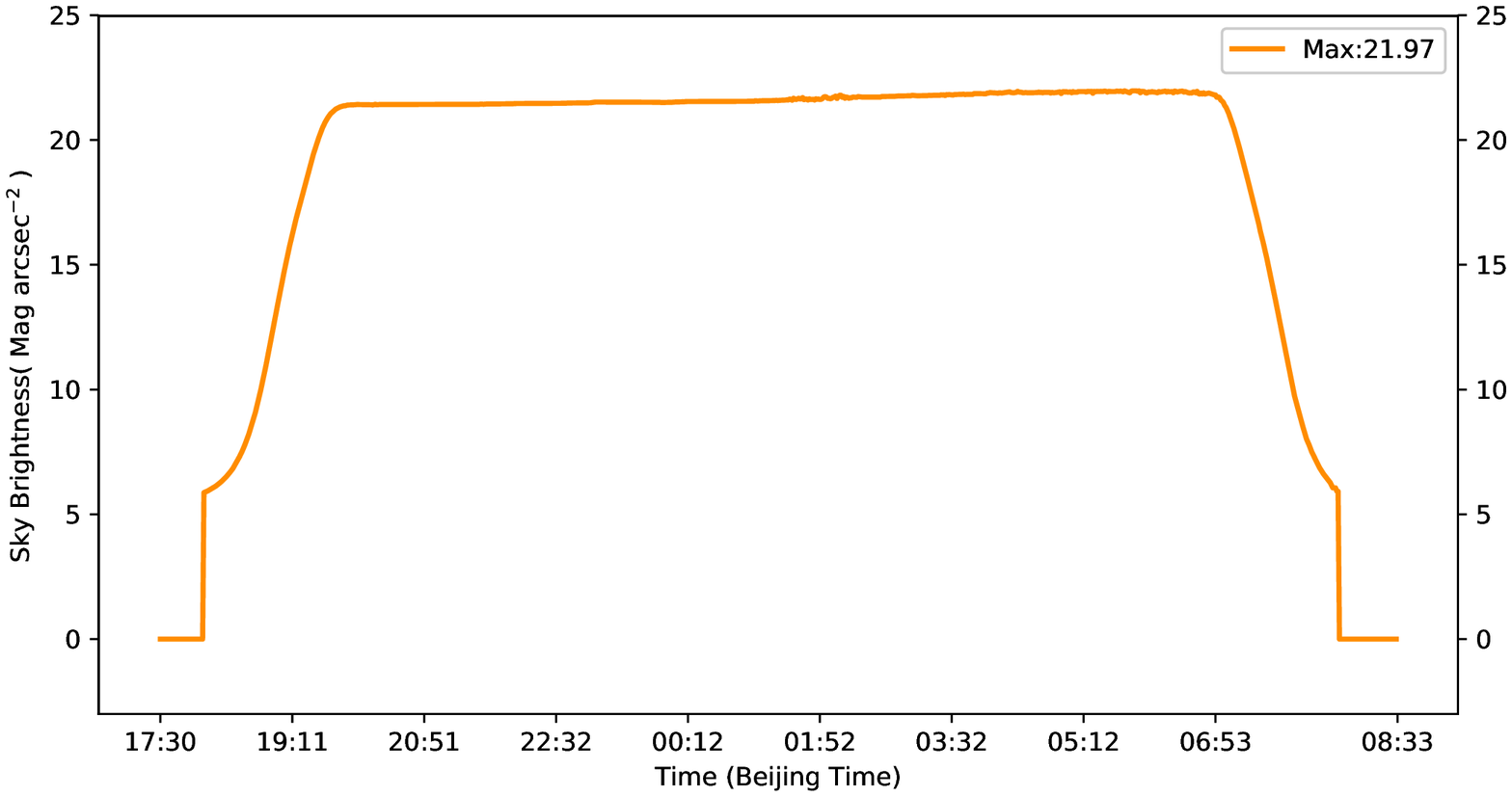}
   \caption{{\small Sky Brightness on 2019-12-25} }
  \label{fig:darkness_max}
  \end{minipage}
  \begin{minipage}[t]{0.5\textwidth}
  \centering
   \includegraphics[width=70mm,height=40mm]{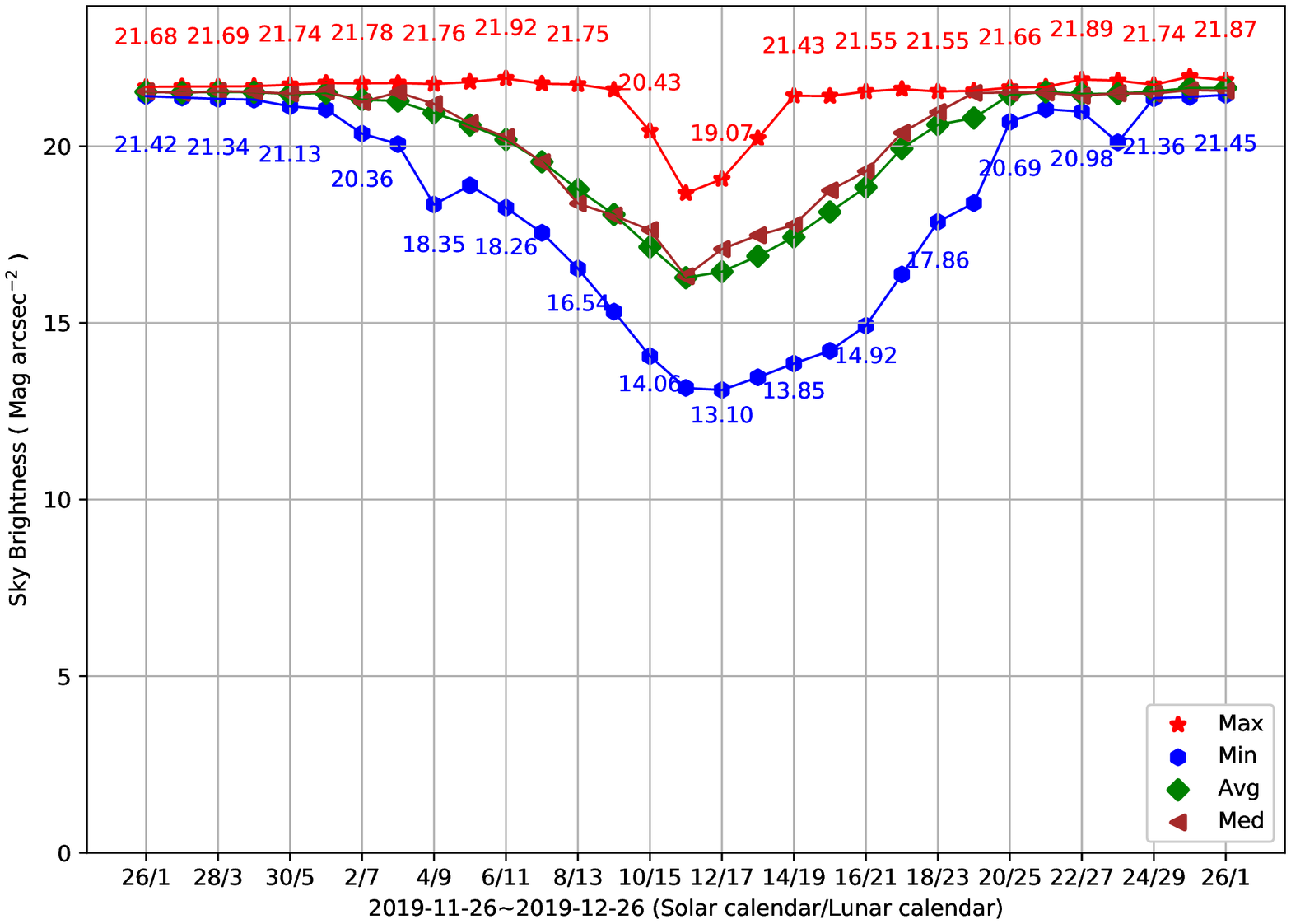}
   \caption{{\small Sky Brightness from 2019-11 to 2019-12}}
  \label{fig:darkness_month}
  \end{minipage}
\end{figure}

The maximum moonless night sky brightness (SQM) value was presented on 2019-12-25 (see Fig. \ref{fig:darkness_max}, and all the values are not subtracted the offset 0.08 from flat glass cover). A lunar month sky brightness variation curve are shown on Figure \ref{fig:darkness_month} (all the data are not subtracted the offset). The minimum brightness exceed 21Mag arcsec$^{-2}$ is shown in ten nights,and it can be considered as moonless nights: $1\sim5$ and $25\sim30$.
The minimum brightness less than 16.5Mag arcsec$^{-2}$ is shown in ten nights, and can be considered as moon nights: $13\sim22$.

Atmospheric extinction is the starlight dimming by the terrestrial atmosphere; the greater the air mass $X(z)$ ( eq.\textasciitilde(\ref{eq:airmass01}) and eq.\textasciitilde(\ref{eq:airmass03}) ) of the starlight traverses through the more it is dimmed. In the astronomical words, this effect is measured using the extinction coefficient $\kappa(\lambda)$, and it must be corrected when we calibrate the instrumental magnitudes. equation\textasciitilde(\ref{Eq:extinction}) is used to get the magnitude $m_{0}(\lambda)$ of the object outside the atmosphere from the magnitude $m(\lambda)$ of the object at the surface of the earth (\citealt{extinction2001}). In 1997, the main extinction coefficient measured at GMG was: $\kappa^{'}_{v}=0.135$, $\kappa^{'}_{b}=0.298$ (\citealt{GMG1999}).


\begin{equation}
\label{Eq:extinction}
    m(\lambda)=m_{0}(\lambda)+\kappa(\lambda)X(z)
\end{equation}

\subsection{Observable Hours and Observable Nights}
\label{subsect:obs_hours}
The observable condition depends on many factors, such as cloud cover, wind speed, relative humidity, air dust, rainfall and light pollution and so on. Lijiang Observatory has no air dust, and almost no light pollution. The Lijiang 2.4-meter telescope could perform normal astronomical observations with wind speed below $15m s^{-1}$ and environmental humidity below 95\%. In Section \ref{subsect:wind}, we have present the wind condition, it is nearly always suitable for LJT's observation, and the heavy wind for a short time only happened one time during the night in the last ten years.

The relative humidity is a worse thing for the Lijiang Observatory, although the humidity inside the dome is 10\% lower than environment. We still have to take measures to reduce the inside humidity by using dehumidifiers and heaters, when the environment humidity is a long time in $95\%\sim 98\% $, even if the heater will bring dome seeing. Anyhow still about $50$ observable hours are lost for very high humidity in a year.

We defined the Observable Hours and Observable Nights, based on the cloud value of the Cloud Sensor (see Sect. \ref{subsubsect:cloud}), as the Lijiang Observatory's Standards:\\
$\bullet$ \textbf{Ideal Hours (IH)}: The time calculated when the sun is below a certain degree, is called Ideal Hours. In GMG station, the certain degree is -13. All the observable hours defined below are subsets of IH. One obvious fact is that: the lower the astronomical site latitude, the greater the IH.  \\
$\bullet$ \textbf{Photometric Hours (PH)}: The cloud is less than and equal to 3 (30\% cloud cover in the sky), and the minimum sustainable time is 1 hour, as a Photometric Time Block (PTB), and the permissible interruption time is 10 minutes in a block.\\
$\bullet$ \textbf{Spectroscopic Hours (SH)}: The cloud is less than and equal to 5 (50\% cloud cover in the sky), and the minimum sustainable time is 1 hour, as a Spectroscopic Time Block (STB), and the permissible interruption time is 10 minutes in a block.\\
$\bullet$ \textbf{Photometric Nights (PN)}: The sum of PTB is exceed 3 hours, as a Photometric Night.\\
$\bullet$ \textbf{Spectroscopic Nights (SN)}: The sum of STB is exceed 3 hours, as a Spectroscopic Night. \\
$\bullet$ \textbf{Clear Nights (CN)}: The sum of PTB is exceed 90\% of IH, as a Clear Night. \\
$\bullet$ \textbf{Ideal Nights (IN)}: The total number of days in a year, 365 for average year and 366 for leap year.

\begin{figure}
   \centering
   \includegraphics[width=0.7\textwidth]{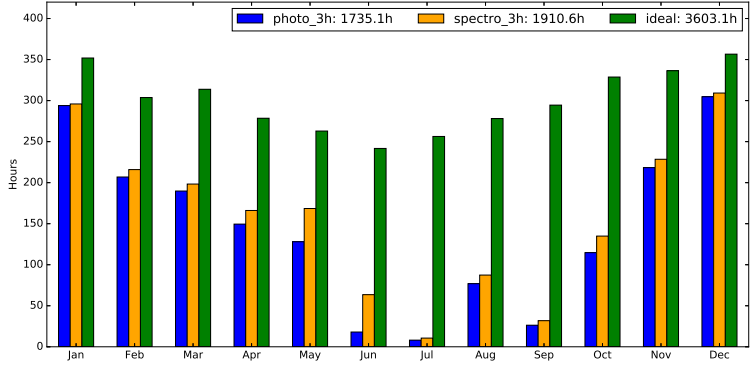}
   \caption{\small Observable Hours in 2019 }
   \label{fig:obshours}
 \end{figure}

\begin{figure}
   \centering
   \includegraphics[width=0.7\textwidth]{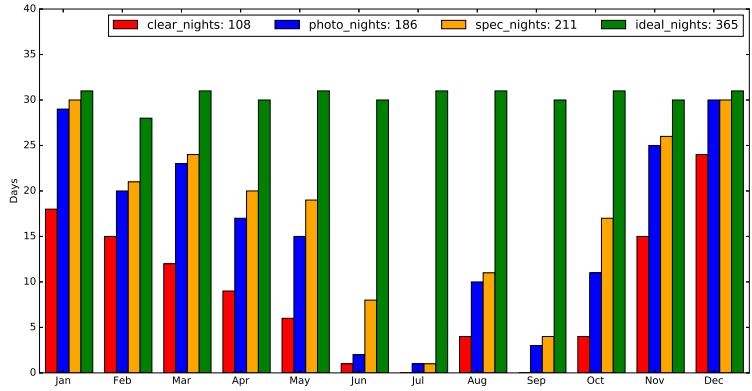}
   \caption{\small Observable Days in 2019 }
   \label{fig:obsdays}
 \end{figure}

We present the three types of observable hours in Figure \ref{fig:obshours}, 48\% of IH is Photometric Hours, and 53\% of IH is Spectroscopic Hours. The summer season months: June, July, August and September is the rain season time, not suitable for astronomical observation. Figure \ref{fig:obsdays} shows the four types of observable nights: Clear Night, Photometric Nights, Spectroscopic Nights and the Ideal Nights. There are 108 clear nights, 186 photometric nights and 211 spectroscopic nights in 2019, reference to historical data, the observable nights of 2019 is less than normal years. The number of photometric nights is about 200 in the normal years.


\subsection{Seeing Data}
\label{subsect:seeing}
Monitoring the atmospheric seeing is an important work, not only for site-survey campaign but also for the conventional observatory. The ADIMM system of ASMS-B collected 346 946 sets of seeing data from February 2019 to January 2020, and the sampling frequency is 12 seconds for the on set. All the the seeing data we analysed in this work, are corrected for the zenith angle. Actually we have saved many raw data in one set, including the arrival angle variances $\sigma_l^2$, $\sigma_t^2$, zenith, Fried parameters before corrected $r_{0l}, r_{0t}$, et al.

\begin{figure}
   \centering
   \includegraphics[width=0.8\textwidth]{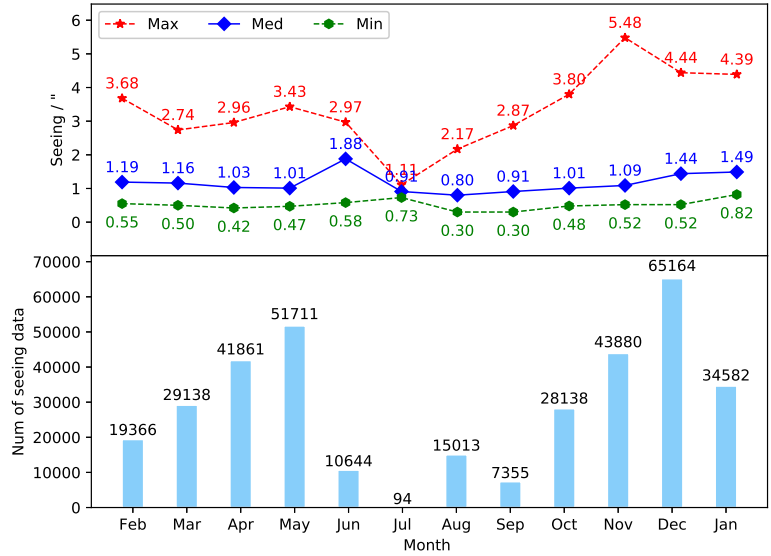}
   \captionsetup{justification=centering}
   \caption{\small Seeing data in 2019-02 $\sim$ 2020-01}
   \label{fig:seeing_2019}
 \end{figure}

\begin{figure}
   \centering
   \includegraphics[width=0.8\textwidth]{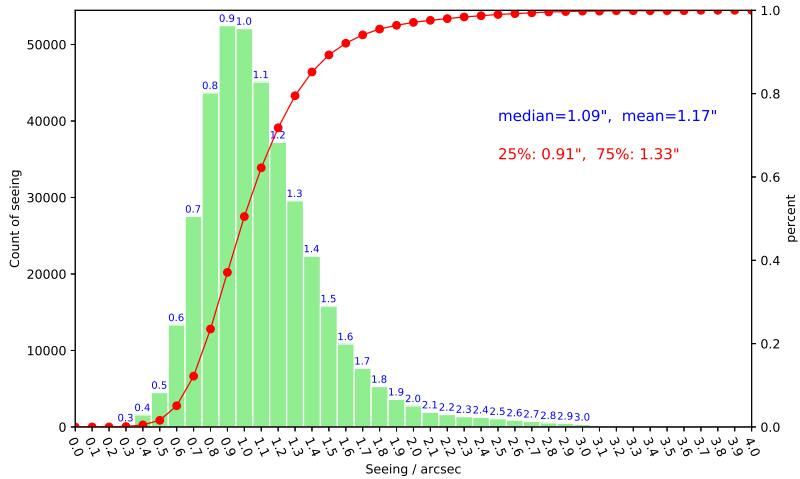}
   \captionsetup{justification=centering}
   \caption{\small Histogram of seeing in 2019-02 $\sim$ 2020-01}
   \label{fig:seeing_2019bar}
 \end{figure}

We present the yearly seeing changes on maximum, minimum and medium, and the counts of $r_0$ in Figure \ref{fig:seeing_2019}. The best seeing condition are in summer months, conversely the seeing conditions in winter months are slightly worse. For some faults happened on ADIMM system's hardware and software, the seeing data are not complete in several month, we upgraded it to a much more robust version since April. The seeing histogram chart in Figure \ref{fig:seeing_2019bar} shows the distribution of seeing data throughout the year, the median is $1.09"$, and the mean is $1.17"$. By percentage, $25\%$ seeing data are below $0.91"$,and $75\%$ seeing data are below $1.33"$. So the Lijiang Observatory is a very good astronomical optical site in China.

\section{Conclusions}
\label{sect:conclusion}
We comprehensively introduced the basic information of Lijiang Observation, and described four parts of Astronomical Site Monitoring System: meteorological station, all sky-information acquisition system, ADIMM system and video surveillance system. We presented the detail design for the cloud sensor, all-sky camera and ADIMM system. Finally, we analyzed the full year data of the GMG station, and get the basic observation conditions for this optical astronomical site.

\begin{acknowledgements}
We are thankful to the work together with BOOTES global network team and TAT global network team. We had a lot of good experiences for building the robotic astronomical site monitoring system. We are thankful for the selfless help from the Xinglong Observatory team for building the first version of the all-sky camera. This work was funded by the National Natural Science Foundation of China (NSFC, under No.11991051, 11203073, 11573067, 11873092 and 11803087), the CAS "Light of West China" Program (No.Y8XB018001).
\end{acknowledgements}

\label{lastpage}

\end{document}